# A network biology-based approach to evaluating the effect of environmental contaminants on human interactome and diseases

Midori Iida[1,2*] and Kazuhiro Takemoto[1*]

[1]Department of Bioscience and Bioinformatics, Kyushu Institute of Technology, Iizuka, Fukuoka 820-8502, Japan

[2]Center for Marine Environmental Studies (CMES), Ehime University, Bunkyo-cho 2-5, Matsuyama, 790-8577, Japan

*Corresponding author's e-mail: iida.midori517@mail.kyutech.jp (M. Iida), takemoto@bio.kyutech.ac.jp (K. Takemoto)

## Abstract

Environmental contaminant exposure can pose significant risks to human health. Therefore, evaluating the impact of this exposure is of great importance; however, it is often difficult because both the molecular mechanism of disease and the mode of action of the contaminants are complex. We used network biology techniques to quantitatively assess the impact of environmental contaminants on the human interactome and diseases with a particular focus on seven major contaminant categories: *persistent organic pollutants (POPs), dioxins, polycyclic aromatic hydrocarbons (PAHs), pesticides, perfluorochemicals (PFCs), metals,* and *pharmaceutical and personal care products (PPCPs)*. We integrated publicly available data on toxicogenomics, the diseasome, protein–protein interactions (PPIs), and gene essentiality and found that a few contaminants were targeted to many genes, and a few genes were targeted by many contaminants. The contaminant targets were hub proteins in the human PPI network, whereas the target proteins in most categories did not contain abundant essential proteins. Generally, contaminant targets and disease-associated proteins were closely associated with the PPI network, and the closeness of the associations depended on the disease type and chemical category. Network biology techniques were used to identify environmental contaminants with broad effects on the human interactome and contaminant-sensitive biomarkers. Moreover, this method enabled us to quantify the relationship between environmental contaminants and human diseases, which was supported by epidemiological and experimental evidence. These methods and findings have facilitated the elucidation of the complex relationship between environmental exposure and adverse health outcomes.

**Abbreviations:** AHR, aryl hydrocarbon receptor; AR, androgen receptor; As, arsenic; BaP, benzo[a]pyrene; CASP, caspase; CRISPR, clustered regularly interspaced short palindromic repeats; CTD, Comparative Toxicogenomics Database; CYP, cytochrome

P450; CXCL8, chemokine (C-X-C motif) ligand 8; DDE, dichlorodiphenyldichloroethylene; DDT, dichlorodiphenyltrichloroethane; ER, estrogen receptor; FDA, Food and Drug Administration; HCB, hexachlorobenzene; IARC, International Agency for Research on Cancer; KS, Kolmogorov-Smirnov; PAHs, polycyclic aromatic hydrocarbons; PARP1, poly (ADP-ribose) polymerase 1; PBBs, polybrominated biphenyls; PBDEs, polybrominated diphenyl ethers; PCB, polychlorinated biphenyls; PCDD, polychlorinated dibenzo-*p*-dioxins; PCDF, polychlorinated dibenzofurans; PFCs, perfluorochemicals; PFOA, perfluorooctanoic acid; PFOS, perfluorooctanesulfonic acid; PFOSF, perfluorooctanesulfonyl fluoride; POPs, persistent organic pollutants; PPAR, peroxisome proliferator-activated receptor; PPCPs, pharmaceutical and personal care products; PPI, protein–protein interaction; PXR, pregnane X receptor; TCDD, 2,3,7,8-tetrachlorodibenzo-*p*-dioxin; TNF, tumor necrosis factor; US EPA, US Environmental Protection Agency.

# 1. Introduction

Humans are exposed to a range of ubiquitous environmental contaminants, which poses a major risk to human health (Gross and Birnbaum, 2017). Studies conducted over the last few decades have revealed the relationships between environmental contaminants, genes, and diseases. For example, dioxins and polycyclic aromatic hydrocarbons (PAHs) cause several types of cancer (Kim et al., 2013; Tavakoly Sany et al., 2015); metallic elements damage multiple organs and cause ailments such as pneumonia, depression, skin lesions, and cancer (Tchounwou et al., 2012); and the persistent exposure to pesticides results in endocrine disruption and polyneuropathy in humans (Hernández et al., 2013). The data on the association between contaminants, genes, and adverse outcomes obtained from previous studies have been curated and deposited in the Comparative Toxicogenomics Database (CTD) (Davis et al., 2017).

The quantification of the association between exposure to individual chemicals and human health outcomes could help guide public health efforts by enabling the intervention in the use of high-risk agents (Braun et al., 2016). However, it is often difficult to evaluate the involvement of contaminants in human disease development because their effects are complex and often indirect (Briggs, 2003). The molecular mechanisms of action of contaminants in inducing human diseases are still poorly understood in the context of the human interactome. From this perspective, network science (Barabási, 2013) and network biology (Barabási and Oltvai, 2004) are useful because networks describe the relationships among elements (nodes) and provide simple and powerful tools for the description and analysis of complicated data sets. In particular, recent computational approaches of network science have successfully used large-scale data on drug-disease associations and disease-gene associations to better understand drug-disease interactions (Goh et al., 2007; Reyes-Palomares et al., 2013; Yıldırım et al., 2007). Yıldırım et al. (2007) constructed drug-target networks that described the relationship between approved/experimental drugs and their target proteins, and systematically evaluated the effects of drugs on the human interactome and diseases using data of essential genes and protein–protein interactions (PPIs). The network-based approach has also been used to obtain comprehensive information on the relationship between contaminants and biological functions. Darabos et al. (2016) investigated the relationships between environmental pollutants and biological pathways to identify candidate biological pathways that might be disrupted by exposure to environmental contaminants. However, the human disease-inducing effects of environmental pollutant are still not fully characterized.

In this study, we quantitatively evaluated the impact of environmental contaminants on the human interactome and diseases using techniques based on network biology and inspired by Yıldırım et al. (2007). We collected datasets on chemical-gene, gene-disease, and chemical-disease associations and then manually classified the chemical compounds into the following seven categories based on the major chemical contaminants, persistent organic pollutants (POPs), dioxins, PAHs, pesticides, perfluorochemicals (PFCs), metals, and pharmaceutical and personal care products (PPCPs). Moreover, we quantified the relationships between chemicals and disease-associated genes in a human PPI network using complex network analysis. In particular, we demonstrated the different effects of the chemicals in each category on human disease development.

## 2. Materials and Methods

*Dataset of chemical-gene interactions*

We obtained the dataset of chemical-gene interactions (chemical targets) from the CTD database (ctdbase.org) (Davis et al., 2017) on July 06, 2017. The dataset contained 12,373 chemical entries. The chemical-target network was represented as a bipartite network in which an edge was drawn between a gene and the chemical compounds that targeted it.

*Classification of chemical compounds*

The chemicals included in this study were selected based on the availability of curated data linked to genes and chemicals, the seriousness of known health effects, and scientific data that suggested exposure risks in humans. We extracted 535 substances that targeted human proteins and classified the substances into the following seven chemical contaminant categories: *POPs, dioxins, PAHs, pesticides, PFCs, metals,* and *PPCPs*, as well as *US Food and Drug Administration (FDA)-approved drugs*. These categories were defined based on the following criteria. However, it should be noted that a chemical compound could be classified into multiple categories.

***POPs*:** The *POP* category was defined based on Annex A of the Stockholm Convention on Persistent Organic Pollutants (UNEP, 2009) and includes organochlorine pesticides (e.g. dieldrin, dichlorodiphenyldichloroethylene (DDE), dichlorodiphenyltrichloroethane (DDT), endosulfan, heptachlor, hexachlorobenzene (HCB), and lindane), polybrominated biphenyls (PBBs), polychlorinated biphenyls (PCBs), and polychlorinated dibenzo-*p*-dioxins and dibenzofuranes (PCDD/Fs) that are commonly known as dioxins, polybrominated diphenyl ethers (PBDE) (tetra-, penta-, hexa-, and hepta-brominated diphenyl ethers), perfluoroalkyl substances (PFOS, its salts, and perfluorooctanesulfonyl fluoride [PFOSF]). Of these chemicals, we extracted 77 available substances from the CTD database.

***Dioxins*:** According to the US Environmental Protection Agency (US EPA) (US EPA, 2004), we categorized the following chemicals as dioxins and dioxin-like compounds: 10 PCDFs, 7 PCDDs, and 12 PCBs. In this study, 12 dioxin-like compounds including 2,3,7,8-tetrachlorodibenzo-*p*-dioxin (TCDD) were available in the CTD database and were classified into the *dioxin* category.

***PAHs*:** PAHs are a large category of chemicals comprising two or more fused aromatic rings. We focused on PAHs in the priority-pollutant lists of the US EPA and the European Food Safety Authority (EFSA) (EFSA, 2008; US EPA, 2014). Eighteen substances including benzo[a]pyrene (BaP) were available in the CTD database and were further analyzed.

***Pesticides***: This category was based on the Updated Tables, January 2017, of the Fourth National Report on Human Exposure to Environmental Chemicals, 2009 (CDC., 2017); in particular, we considered the substances assigned to the following categories as pesticides: insect repellents and metabolites, carbamate pesticide metabolites, organochlorine pesticide metabolites, organophosphorus insecticides, pyrethroid metabolites, and organochlorine pesticides and metabolites. Finally, 18 substances including DDE and DDT were available in the CTD database and were included in our analysis.

***PFCs***: PFCs were defined as chemical substances assigned to the perfluoroalkyl polyfluoroalkyl categories of the Fourth Report (CDC., 2017). Eight available substances including perfluorooctanoic acid (PFOA) and PFOS were identified in the CTD database and were included in our analysis.

***Metals***: The substances assigned to the metal and metalloid category in the Fourth Report (CDC., 2017) and the heavy metal category in the CTD database (Davis et al., 2017) were included in this category. A total of 74 substances such as arsenic (As), copper (Cu), and zinc (Zn) from the CTD database were analyzed in this study.

***PPCPs***: Despite the absence of a consensus on the definition of PPCPs, in this study they were defined as substances assigned as personal care and consumer product chemicals and metabolites in the CDC Updated Tables of the Fourth Report (CDC., 2017) and several previous studies (Batt et al., 2016; Howard and Muir, 2011; Tanoue et al., 2015). We analyzed 18 substances that included acetaminophen and ibuprofen, which were available in the CTD database.

***FDA-approved drugs***: For comparison with a previous study (Yıldırım et al., 2007), we also considered approved drugs. We identified FDA-approved drugs from the DrugBank database (drugbank.ca) (Law et al., 2014) on May 1, 2017. We included 331 substances such as fluorouracil that were available in the CTD database in the analysis.

*Gene-disease and chemical-disease association datasets*

We obtained datasets of gene-disease and chemical-disease associations from the CTD database (Davis et al., 2017) on July 06, 2017. The dataset of gene-disease associations consisted of 54,876,955 associations between 5,792 diseases (including disorders) and 42,710 disease-associated genes. The dataset of chemical-disease associations contained 5,080,629 associations between 5,814 diseases and 14,978 disease-associated chemicals. We extracted only the associations supported by therapeutic or marker/mechanism in the *Direct Evidence* field as direct evidence. Finally, the datasets of the gene-disease and chemical-disease associations consisted of 26,742 associations between 4,400 diseases and 7,352 disease-associated genes, and 93,464 associations between 3,152 diseases and 9,300 disease-associated chemicals, respectively.

*Human protein–protein interaction network*

We obtained physiological human PPI data from a previous study (Yıldırım et al., 2007). We used datasets of PPIs from three high-quality systematic yeast two-hybrid experiments and the published literature using manual curation (Goh et al., 2007; Rolland et al., 2014; Rual et al., 2005; Stelzl et al., 2005). We integrated the datasets to obtain the PPI data that consisted of 37,946 non-self-interacting and non-redundant interactions between 10,367 proteins (Table S1). The PPI network was presented as a binary unipartite network in which nodes and edges represented the proteins and physiological PPIs, respectively. We mapped the chemical target and disease-associated proteins onto the PPI network.

*Essential human genes*

To compare the node degree in essential genes and chemical target proteins in the human PPI and determine whether the chemical target proteins are enriched in essential genes, we obtained a dataset of essential genes identified using the clustered regularly interspaced short palindromic repeats (CRISPR) system from a previous study (Blomen et al., 2015). The degree of differences was compared using the Wilcoxon rank-sum test, and the enrichments were statistically evaluated using Fisher's exact test.

*Evaluating effects of chemical compounds on human diseases using PPI network*

According to the methods of a previous study (Yıldırım et al., 2007), we evaluated the impact of chemicals on human diseases using a PPI network. The impact was evaluated based on the distance (shortest path length) between the chemical target and disease-associated proteins in the PPI network compared to random controls. The disease-associated genes were classified into 34 disease categories according to the MEDIC-Slim classification terms of the CTD database (Davis et al., 2013). Italic font was used to indicate the MEDIC-Slim categories to distinguish the terms from disease names. In this study, we only focused on 14 categories (*cancer, cardiovascular disease, congenital abnormality, digestive system disease, endocrine system disease, immune system disease, mental disorder, metabolic disease, musculoskeletal disease, nervous system disease, pregnancy complication, respiratory tract disease, signs and symptoms,* and *skin disease*) that had sample sizes sufficient for statistical tests. First, we calculated the minimum value ($D_{\text{real}}$) of the distance between the chemical targets and the disease-associated proteins. Then, we calculated the distribution of the minimum distance for each disease in the category. To generate the controls, we randomly selected proteins with the same number of chemical target proteins from the PPI network 500 times to control the chemical targets. We kept the disease-associated proteins constant and calculated the minimum values ($D_{\text{rand}}$) for the distance between randomly selected targets and disease-associated proteins. The effect of chemicals on human disease was based on the difference between the actual data and the random controls. We evaluated the difference between the distribution of $D_{\text{real}}$ and $D_{\text{rand}}$ using the Kolmogorov-Smirnov (KS) test for each chemical category. Moreover, the differences in the mean values between $D_{\text{real}}$ and $D_{\text{rand}}$ were evaluated using Welch's two-sample *t*-test, which was quantified using the log2 ratio, defined as $\log_2(\langle D_{\text{real}} \rangle / \langle D_{\text{rand}} \rangle)$. Furthermore, $\langle x \rangle$ indicates the mean value of *x*. A smaller log2 ratio was indicative of a more direct effect of the chemicals on the disease; a log2 ratio ≥ 0 indicated an indirect effect on the disease. When the log2 ratio was < 0 and the mean values between $D_{\text{real}}$ and $D_{\text{rand}}$ were not significantly different, it was also considered to indicate an indirect effect.

*Statistical analyses and network analyses*

The statistical and network analyses were performed using the R packages *stats* (version 3.3.2), *exactRankTests* (version 0.8.29), and *igraph* (version 1.1.2) in R software (version 3.3.2; www.r-project.org). The node degree (i.e., the number of edges per node) was calculated using the degree function in the *igraph* package to evaluate the characteristics of each node. In this context, we also calculated the following nodal measures (Freeman, 1978; Takemoto and Oosawa, 2012): closeness centrality, betweenness centrality, and clustering coefficients using *closeness*, *betweenness*, and *transitivity* functions in the *igraph* package, respectively. The closeness centrality is based on the shortest path length between nodes. A node with a large closeness centrality indicates that the average path length between that and the other nodes is relatively short. The betweenness centrality describes a walker's movement from one node to another via the shortest path; therefore, nodes with a large number of visits by the walker shows high betweenness centrality. The clustering coefficient of a node characterizes the edge density among neighbors of the node. The shortest path length of a given node pair was obtained using the *distances* function in the *igraph* package. The differences in the median degree were evaluated based on the Wilcoxon rank-sum test; specifically, the *wilcox.exact* function in the *exactRankTests* library. Fisher's exact test, the KS test, and Welch's two-sample *t*-test were computed using the functions *fisher.test*, *ks.test*, and *t.test* in the *stats* library, respectively.

## 3. Results

*Chemical-target network*

The extracted CTD dataset included 12,304 interactions for 3,326 genes and 535 chemical compounds including 77 *POPs*, 12 *Dioxins*, 18 *PAHs*, 18 *Pesticides*, 8 *PFCs*, 74 *Metals*, 18 *PPCPs,* and 331 *FDA-approved drugs* (Table S2), which were used to construct the chemical-target (i.e., chemical-gene interaction) network (Fig. 1). To characterize the connectivity tendencies, the frequency node degree distributions in the chemical-target networks were investigated. In bipartite networks, we considered the distributions of chemical and target nodes. The node degrees of a chemical compound and gene (chemical target) correspond to the number of genes targeted by the chemical ($n_c$) and the number of chemicals targeting to the gene ($n_g$), respectively. The frequency distributions of $n_c$ and $n_g$ were fat-tailed (Fig. 2), which indicated that most chemicals targeted a few genes; however, a few chemicals had many targets. The top five chemicals ranked by $n_c$ for each chemical category are summarized in Table S4. Most notably, we identified discriminative hub chemicals for each chemical category. For example, the highest node degree chemicals were endosulfan, TCDD, BaP, DDE, PFOA, copper, acetaminophen, and fluorouracil in the *POPs*, *dioxins*, *PAHs*, *pesticides*, *PFCs*, *metals*, *PPCPs*, and *FDA-approved drugs* categories, respectively (Fig. 2a and Table S4). The fat-tailed degree distributions for chemical-target nodes (Fig. 2b) indicated that most genes were targeted by a few chemicals; however, a few genes were targeted by many chemicals. The top five chemical targets ranked by $n_g$ for each chemical category are summarized in Table S5. The hub target proteins tended to be common among the chemical categories; for example, sex hormone receptors (e.g., androgen receptor [AR] and estrogen receptors [ERs]) were targeted by a number of chemicals in different chemical categories.

*Chemical targets in the human protein–protein interaction network*

We examined the global relationships between chemical targets in the human PPI network and identified 231, 208, 276, 1000, 144, 79, 281, 1783, 2212, and 1061 chemical-target proteins in the PPI networks for *POPs*, *dioxins*, *PAHs*, *pesticide*, *PFCs*, *metals*, *PPCPs,* and *FDA-approved drugs* categories, respectively (Fig. S1). For comparison, the node degrees of any and essential proteins were also considered (Fig. 3). We then investigated the node degree (i.e., the number of interacting partners) in the PPI network for each chemical category (Fig. 3) and found that the chemical-target proteins had more interacting partners than any and essential proteins did. The median node degrees of the chemical targets were 76.8%–382.9% larger than that of any proteins in the PPI network ($p < 10^{-8}$ using the Wilcoxon rank-sum test); moreover, the median node degrees of the chemical targets were higher than that the essential proteins ($p < 10^{-6}$ using the Wilcoxon rank-sum test). Furthermore, we investigated the other nodal measures, closeness centrality, betweenness centrality, and clustering coefficients, and found similar trends in the degree centrality (Fig. S2). In particular, the proteins that were targeted by chemicals, including those encoded by essential genes (*ESGs*), had higher median closeness centrality than any proteins did (Fig. S2a; $p < 10^{-4}$ using the Wilcoxon rank-sum test). Moreover, the centralities of *POPs*, *dioxins*, *PAHs*, *pesticides*, and *multiple* targeted proteins were higher than that of *ESGs* (Fig. S2a; $p < 10^{-5}$ using the Wilcoxon rank-sum test). The median betweenness centrality of chemical targets was significantly higher than that of any proteins and *ESGs* (Fig. S2b; $p < 10^{-5}$ with the Wilcoxon rank-sum test). *POPs*, *dioxins*, *PAHs*, *pesticides* and *multiple* targeted proteins showed a higher clustering coefficient than that of all proteins and *ESGs* (Fig. S2c; $p < 10^{-5}$ using the Wilcoxon rank-sum test). These results suggest that contaminant targets are hub proteins and have a major effect on the human interactome.

*Chemical targets and gene essentiality*

The node degree is positively correlated with gene essentiality in the human PPI network (Jeong et al., 2001). Since the chemical targets had numerous interacting partners in the PPI network (Fig. 3), we hypothesized that chemical targets would be enriched in essential protein datasets. The association analysis showed that the *PAHs*, *metals*, and *FDA-approved drug* targets were enriched in human essential proteins; however, no positive associations were observed between the chemical targets in the other categories and essential genes. Instead, *POPs*, *dioxins*, and *multiple* targets were depleted with respect to essential genes (Fig. 4). These results suggest that chemical-target proteins did not necessarily represent an abundance of essential genes, although they had larger interactions than essential genes did.

*Cellular network-based evaluation of effects of chemical targets on human diseases*

We investigated the relationship (distance) between chemical targets and disease-associated proteins in the human PPI network to evaluate how chemicals in each specific chemical category would directly affect human diseases. From the dataset of chemical-disease associations, we extracted chemical targets and disease-associated proteins from the PPI network and obtained 65,941 chemical-disease associations. We identified a remarkable enrichment in all chemical categories compared with the random controls at shorter than longer distance regions (Fig. 5). In all chemical categories, the distribution of the distance between the actual data and random controls was significantly different (Fig. 5; $p < 0.01$ using the KS test). Therefore, the distances between disease-associated and chemical-target proteins were shorter than those of the random control, suggesting the direct effects of these chemicals on human diseases.

To further investigate the effect of these chemicals on human diseases, we focused on 14 disease categories and compared the distance between the chemical targets and disease-associated proteins for each chemical category (Figs. 6, S3, and Table S6). Most disease categories had negative log2 values, which indicated that the chemical target and disease-associated proteins were more closely located than they were in the random control in the PPI network. A small log2 ratio indicated direct effects of the chemicals on human diseases. Specifically, all chemical categories directly influenced *cancer* and *digestive system disease* including colonic neoplasms, fatty liver, and hepatitis. However, the effect of chemicals on human diseases occasionally differed between the chemical categories. For example, *POPs*, *dioxins*, *metals*, *PPCPs*, and *FDA-approved drugs* had direct effects on *endocrine system disease*, *immune system disease, nervous system disease,* and *respiratory tract disease* categories, whereas *dioxins* indirectly affected diseases in the *mental disorder* category. Most distances between *PFCs* targets and disease-associated proteins were similar to the random control, which suggests that chemicals in the *PFCs* category indirectly affected human diseases. Similarly, *PAHs* indirectly affected proteins in the *endocrine system disease, immune system disease,* and *nervous system disease* categories. *Pesticides* indirectly affected *immune system disease, nervous system disease, respiratory tract disease*, and *signs and symptoms* (e.g., fever, headache, and chills).

## 4. Discussion

In this study, we focused on environmental contaminants and evaluated their effects on the human interactome and diseases using network biology techniques. We investigated the chemical-target network and found that the degree of chemical distributions and their targets were fat-tailed (Fig. 2). Similar distributions were also observed in the drug-target network described in this (Fig. 2 "approved drugs") and previous (Yıldırım et al., 2007) studies. Yıldırım et al. (2007) investigated the drug-target network formed by interactions between FDA-approved drugs and their target proteins and demonstrated that most drugs targeted only a few proteins. The contaminant-target network also suggested that most genes were targeted by a few chemicals, while a few genes were targeted by many chemicals. The node degree of chemicals in the chemical-target network could reflect their ecotoxicological significance. For example, TCDD and BaP were significant hubs in the network (Fig. 2a). Historically, these chemicals have been used as representative substances of their chemical class for exposure experiments (Fracchiolla et al., 2016; Kim et al., 2013).

The proteins targeted by chemicals in different categories might have important roles in evaluating the combined effects of environmental pollutants. For example, the AR and ERs were conserved hub genes among the *POPs*, *dioxins*, *PAHs,* and *pesticides* categories, despite the differences in their chemical properties. Indeed, ERs are widely used to evaluate combined or mixed effects of xenoestrogens (e.g., *POPs, dioxins, PAHs, and pesticides*) such as endocrine disruption (Silins and Högberg, 2011). Moreover, the node degree of chemical targets is useful in identifying biomarkers of chemical exposure. For example, aryl hydrocarbon receptor (AHR), cytochrome P450 1A1 (CYP1A1), and CYP1A2 were hub proteins in the *dioxins* and *PAHs* categories (Fig. 2b and Table S5). CYP1A1 is a widely used biomarker of the exposure of contaminants that bind to the AHR (Petrulis et al., 2001). Pregnane X receptor (PXR), CYP3A4, and CYP2B6 were hub targets in the *pesticide* category, which is consistent with the activation of PXR by organochlorine pesticides that induces the transcription of *CYP3A4* and *CYP2B6* (Bertilsson et al., 1998; Coumoul et al., 2002). Similarly, peroxisome proliferator-activated receptor α (PPARα) and PPARγ were hub targets in *PFCs*, and are activated by PCFs (Abbott, 2009; Rosen et al., 2017; Wolf et al., 2014). Chemokine (C-X-C motif) ligand 8 (CXCL8), tumor necrosis factor (TNF), and caspase-3 (CASP3) were hub proteins in the *metal* category. These proteins are reported as targets for metals (Carter et al., 1997; Khan et al., 2011; Kim et al., 2008), and they play important roles in tumor progression (Hamed et al., 2012; Knüpfer and Preiß, 2007). Because AR and poly (ADP-ribose) polymerase 1 (PARP1), which is associated with DNA repair and cellular differentiation (Tao et al., 2009; Vinggaard et al., 2000), were hub targets in the *PAH* category, we hypothesized that AR and PARP1 are important biomarkers for PAHs. However, as their roles in toxicity are not well described, further examination is needed to verify this hypothesis.

The contaminant target proteins had more interacting partners than any other proteins in the human PPI network did (Fig. 3). This trend was also observed for drug targets (Yıldırım et al., 2007). These results indicate that contaminant targets are hub proteins with a major effect on the human interactome because hubs are related to network robustness. Cancer treatments could be interpreted as targeted attacks on hub proteins such as p53 (Lane et al., 2010), and the fragility of networks is determined by these targeted attacks on hubs (Albert et al., 2000). Thus, the node degree of a chemical target in the PPI networks may be useful in estimating the effect of a chemical on the human interactome. In this context, other nodal measurements such as centrality measures may also be informative parameters to estimate the effect of chemicals on the human interactome. The analyses based on closeness centrality (Fig. S2a) and betweenness centrality (Fig. S2b) showed that the contaminant target proteins were more centric than any of the proteins in the human PPI network were. Closeness centrality was used to measure how fast the flow of information would be through a specific node to other nodes. Thus, this result suggests that chemical stimuli such as *POPs*, *dioxins*, *PAHs*, *pesticides*, and *multiple* protein categories rapidly spread in the human interactome compared to other chemicals such as *PFCs* and *metals*. A high betweenness centrality relies on communication paths and could control information flow; therefore, our results also indicate that pollutants and drugs considerably affect the human interactome. In this study, we found that the contaminant target proteins were more clustered than any proteins in the human PPI network were (Fig S2). The clustering coefficient of a node (protein) characterizes the edge density among node neighbors and, thus, it is useful for identifying protein complexes. *POPs-*, *dioxins-*, *PAHs-*, *pesticides-* and *multiple-* targeted proteins have tightly clustered interactions, and are expected to form large complexes.

Association analyses between target proteins and essential proteins (Fig. 3) could be useful for estimating the effect of chemicals on humans, which are difficult to evaluate. *POP*, *dioxins*, and *pesticide* targets have many interacting partners; however, essential human proteins were not abundant among these targets (Fig. 4). This indicated that these target proteins have many interactions but are not essential for life. In contrast, the *PAH*, *metal*, and *FDA-approved drug* targets had a high node degree, and essential proteins were abundant among their targets, which indicates that these chemicals might have more severe effects on organisms than the other chemical categories do. PPCPs are also FDA-approved drugs; thus, they are expected to have the same interaction patterns as other FDA-approved drugs. However, unlike *FDA-approved drug* targets, *PPCP* targets were not abundant in essential protein categories. Therefore, we would expect PPCPs to have broad effects on the human interactome and lower lethality than *FDA-approved drug*s.

The distance between the chemical targets and disease-associated proteins estimates the number of molecular steps between a chemical target and the corresponding disease-associated protein in the PPI network (Yıldırım et al., 2007). A shorter distance between chemical targets and disease-associated proteins indicates that the chemicals exert a more direct effect on diseases than a longer distance does. The network-based evaluation of these associations is useful for quantifying the effect of chemicals on human disease. Yıldırım et al. (2007) demonstrated that FDA-approved drugs had a higher enrichment of proteins in the region of the lower distances than the random control did. A high enrichment in the lower distances was also observed in this study (Fig. 5). Moreover, the distribution of the distances of all the contaminant categories showed clear enrichments at shorter distances similar to the previous study (Fig. 5) (Yıldırım et al., 2007). These results indicate that environmental contaminants are likely direct causes of human diseases.

Finally, we evaluated the effect of chemical targets on a specific disease group (Fig. 6 and Table S6) and found that most contaminants directly affected human diseases. *Cancer* and *digestive system disease*, including liver disease, were directly affected by all environmental contaminant and chemical drug categories (Fig. 6). This result is supported by a large body of evidence collected on the relationship between the exposure to these contaminants and the elevated risk of cancer development (Bahadar et al., 2014; Grandjean and Clapp, 2014; Kim et al., 2013; Tavakoly Sany et al., 2015; Tchounwou et al., 2012). For example, the International Agency for Research on Cancer (IARC) has reported that BaP, TCDD, and metal compounds should be classified as category one human carcinogens (IARC, 2010). The liver is prone to xenobiotic-induced injury because it plays an important role in xenobiotic metabolism (Sturgill and Lambert, 1997); and POPs mixtures, PAHs, pesticides, PFCs, and metals are known to have negative effects on liver health in humans and experimental animals (Carlin et al., 2016; Deierlein et al., 2017; Lindstrom et al., 2011; Min et al., 2015; VoPham et al., 2017).

*POPs*, *dioxins*, *pesticides*, and *metals* had direct effects on members of the *endocrine system disease* category (Fig. 6). This may be because these environmental contaminants include major endocrine disrupting chemicals that perturb the normal synthesis, secretion, transportation, binding, and metabolism of natural hormones (Annamalai and Namasivayam, 2015). Generally, endocrine disruptors can exert their effects via two pathways: either directly on hormone-receptor complexes or directly on the specific proteins involved in the control of hormone delivery to the right place at the right time (WHO and UNEP, 2013). Since *PAHs* and *PFCs* showed indirect effects on the *endocrine system disease* category in this study (Fig. 6), the reported effects of these chemicals on components of the endocrine system could be induced by hormone-related proteins.

*POPs*, *dioxins*, and *metals* directly affected the *immune system disease* group (Fig. 6 and Table S6). In fact, chemicals in these categories are known to induce immunological disorders (Colborn et al., 1993; Gascon et al., 2013; Tchounwou et al., 2012). In particular, AHR, a dioxin receptor, modulates the immune system through regulatory T-cell (Treg) expansion (Tavakoly Sany et al., 2015; Veldhoen et al., 2008). This is consistent with the lowest log2 ratio calculated for the *dioxin* category in the *immune system disease* group. In contrast, the effects of *PAHs*, *pesticides*, and *PFCs* on the *immune system disease* group were indirect, as the immunotoxic effects of PAHs are not directly caused by the parent compounds (Abdel-Shafy and Mansour, 2016). Moreover, epidemiological studies have shown the low immunotoxic risk of pesticides and PFOA (a type of PFC) (Corsini et al., 2013; Steenland et al., 2010). Although these epidemiological studies support our findings, experimental studies indicate a direct association between immune system disorders and pesticides/PFOA (Corsini et al., 2013; Steenland et al., 2010). Further investigations would be required to clarify the effects of pesticides and PFCs on the human immune system.

In addition to the general toxicities of PFCs (Grandjean and Clapp, 2014), our results suggest a direct effect of PFCs on nervous system diseases. This result is consistent with the negative association between blood PFC levels and impulsivity/short-term memory impairment observed in children (Gallo et al., 2013; Gump et al., 2011). Furthermore, increased PFOA concentrations were associated with elevated cholesterol and uric acid levels in humans, which may cause cerebrovascular diseases (Miralles-Marco and Harrad, 2015) that are categorized in the *nervous system disease* group. The chemical-target network analysis showed that PFCs were related to PPARα and PPARγ (Fig. 2b). PFCs possibly cause cerebrovascular diseases through PPARs and, therefore, the PPAR family (α, γ, and β/δ) could be a therapeutic target for cerebrovascular diseases (Nicolakakis and Hamel, 2010).

Previous studies suggest that pesticide exposure can increase the risk of nervous system diseases such as Parkinson's disease (PD) and amyotrophic lateral sclerosis (ALS) in humans (Nandipati and Litvan, 2016; Sánchez-Santed et al., 2016). However, in this study, we did not find a direct effect of *pesticides* on the *nervous system disease* category (Fig. 6). This may have been because many pesticides indirectly contribute to nervous system diseases (Nandipati and Litvan, 2016), but some pesticides such as organochlorine pesticides, could directly contribute to nervous system disease development (Sánchez-Santed et al., 2016).

Our analysis identified a significant direct effect of *PAHs* but not *pesticides* and *PFCs* on *respiratory tract disease*. This was supported by several pieces of evidence, including those suggesting that PAHs are formed mainly by organic fuel combustion and released into the atmosphere (Baek et al., 1991), where they can contribute to risk factors for respiratory cancer and childhood asthma (Bosetti et al., 2007; Karimi et al., 2015; Rota et al., 2014). A meta-analysis study concluded that the relationship between pesticide exposure and respiratory health is controversial (Mamane et al., 2015), and several other studies suggest that occupational exposure to pesticides increases the risks of developing asthma (Fareed et al., 2013; Ye et al., 2013). However, there is little epidemiological evidence to support the relationship between PFC exposure and asthma in humans (Dong et al., 2013), although a study in mice indicated a possible association (Fairley et al., 2007).

Our analysis suggests that PPCPs directly affect mental disorders, nervous system diseases, and signs and symptoms such as fever, headache and chills, heart murmurs, and motion sickness. Indeed, the *PPCPs* category includes antidepressants such as sertraline, fluoxetine, and amitriptyline, which are used for the treatment of nervous system diseases and mental disorders, and antipyretic analgesics such as acetaminophen and ibuprofen for the treatment of fever and headache. Currently, the human health risk from exposure to PPCPs has indicated little cause for concern; however, the effect of PPCPs on human health may be significant in the future because of the combination of increased PPCP use and potential drought increase PPCP concentrations in wastewaters over the next few decades (Cizmas et al., 2015).

While our results provide new data in evaluating the effect of environmental contaminants on human PPI and diseases, it is important to also understand the limitations of the methodology. First, the effective concentrations and adverse outcomes of chemicals depend on the route of exposure (Damalas and Eleftherohorinos, 2011); however, data on the route of exposure was not available at the time of our study. Second, we selected chemicals that have been linked to gene expression changes, human health risks, and have suggested exposure risks to humans. Thus, chemical compounds that have no information on related genes, adverse outcome/disease, or exposure status were excluded from our analysis. Third, we evaluated the relationship between environmental contaminants and human diseases, although the cumulative risk of chemical mixtures was not investigated. To avoid these limitations in the future, larger-scale and more normalized databases should be constructed through data sharing. In addition to these experimental considerations, additional quantitative approaches could improve the prospective mixture risk assessment. Nonetheless, our findings provide a novel tool for identifying biomarkers to evaluate toxic effects of several chemical categories and evaluating the strength of relationships between relevant environmental contaminants and human diseases.

## 5. Conclusions

We quantitatively evaluated the effect of environmental contaminants on the human interactome and diseases using a network biology-based approach. The chemical-target network identified proteins with important roles in the human interactome and its sensitivity to multiple chemicals. Although chemical-target proteins had more interacting partners than essential genes did, they do not necessarily represent an abundance of essential genes. Our analyses suggest that environmental contaminants may directly affect disease-associated proteins and, consequently, cause specific diseases. However, the degree of the effect of each environmental contaminant on human diseases depends on the disease type and chemical category. Epidemiological and experimental evidence support our findings, and the methods we used could enhance our understanding of the relationship between environmental exposure and adverse human health outcomes.

## Acknowledgments

This study was supported by a Grant-in-Aid for Young Scientists (A) from the Japan Society for the Promotion of Science (No. 17H04703). The authors thank Dr. Muhammed A. Yıldırım for his helpful comments on human PPI networks and are much obliged to Dr. Rumi Tanoue for her useful comments on the classification of PPCPs.

## References

Abbott, B.D., 2009. Review of the expression of peroxisome proliferator-activated receptors alpha (PPARα), beta (PPARβ), and gamma (PPARγ) in rodent and human development. Reprod. Toxicol. 27, 246–257. https://doi.org/10.1016/j.reprotox.2008.10.001

Abdel-Shafy, H.I., Mansour, M.S.M., 2016. A review on polycyclic aromatic hydrocarbons: Source, environmental impact, effect on human health and remediation. Egypt. J. Pet. 25, 107–123. https://doi.org/10.1016/J.EJPE.2015.03.011

Albert, R., Jeong, H., Barabási, A.-L., 2000. Error and attack tolerance of complex networks. Nature 406, 378–382. https://doi.org/10.1038/35019019

Annamalai, J., Namasivayam, V., 2015. Endocrine disrupting chemicals in the atmosphere: Their effects on humans and wildlife. Environ. Int. 76, 78–97. https://doi.org/10.1016/j.envint.2014.12.006

Baek, S.O., Field, R.A., Goldstone, M.E., Kirk, P.W., Lester, J.N., Perry, R., 1991. A review of atmospheric polycyclic aromatic hydrocarbons: Sources, fate and behavior. Water. Air. Soil Pollut. 60, 279–300. https://doi.org/10.1007/BF00282628

Bahadar, H., Mostafalou, S., Abdollahi, M., 2014. Current understandings and perspectives on non-cancer health effects of benzene: A global concern. Toxicol. Appl. Pharmacol. 276, 83–94. https://doi.org/10.1016/j.taap.2014.02.012

Barabási, A.-L., 2013. Network science. Philos. Trans. A. Math. Phys. Eng. Sci. 371, 20120375. https://doi.org/10.1098/rsta.2012.0375

Barabási, A.-L., Oltvai, Z.N., 2004. Network biology: understanding the cell's functional organization. Nat. Rev. Genet. 5, 101–113. https://doi.org/10.1038/nrg1272

Batt, A.L., Kincaid, T.M., Kostich, M.S., Lazorchak, J.M., Olsen, A.R., 2016. Evaluating the extent of pharmaceuticals in surface waters of the United States using a National-scale Rivers and Streams Assessment survey. Environ. Toxicol. Chem. 35, 874–81. https://doi.org/10.1002/etc.3161

Bertilsson, G., Heidrich, J., Svensson, K., Asman, M., Jendeberg, L., Sydow-Bäckman, M., Ohlsson, R., Postlind, H., Blomquist, P., Berkenstam, A., 1998. Identification of a human nuclear receptor defines a new signaling pathway for CYP3A induction. Proc. Natl. Acad. Sci. U. S. A. 95, 12208–13.

Blomen, V.A., Májek, P., Jae, L.T., Bigenzahn, J.W., Nieuwenhuis, J., Staring, J., Sacco, R., van Diemen, F.R., Olk, N., Stukalov, A., Marceau, C., Janssen, H., Carette, J.E., Bennett, K.L., Colinge, J., Superti-Furga, G., Brummelkamp, T.R., 2015. Gene essentiality and synthetic lethality in haploid human cells. Science 350, 1092–6. https://doi.org/10.1126/science.aac7557

Bosetti, C., Boffetta, P., La Vecchia, C., 2007. Occupational exposures to polycyclic aromatic hydrocarbons, and respiratory and urinary tract cancers: a quantitative review to 2005. Ann. Oncol. Off. J. Eur. Soc. Med. Oncol. 18, 431–46. https://doi.org/10.1093/annonc/mdl172

Braun, J.M., Gennings, C., Hauser, R., Webster, T.F., 2016. What Can Epidemiological Studies Tell Us about the Impact of Chemical Mixtures on Human Health? Environ. Health Perspect. 124, A6-9. https://doi.org/10.1289/ehp.1510569

Briggs, D., 2003. Environmental pollution and the global burden of disease. Br. Med. Bull. 68, 1–24. https://doi.org/10.1093/bmb/ldg019

Carlin, D.J., Naujokas, M.F., Bradham, K.D., Cowden, J., Heacock, M., Henry, H.F., Lee, J.S., Thomas, D.J., Thompson, C., Tokar, E.J., Waalkes, M.P., Birnbaum, L.S., Suk, W.A., 2016. Arsenic and environmental health: State of the science and future research opportunities. Environ. Health Perspect. 124, 890–899. https://doi.org/10.1289/ehp.1510209

Carter, J.D., Ghio, A.J., Samet, J.M., Devlin, R.B., 1997. Cytokine production by human airway epithelial cells after exposure to an air pollution particle is metal-dependent. Toxicol. Appl. Pharmacol. 146, 180–8. https://doi.org/10.1006/taap.1997.8254

CDC., 2017. Fourth Report on Human Exposure to Environmental Chemicals, 2009. Atlanta, GA.

Cizmas, L., Sharma, V.K., Gray, C.M., McDonald, T.J., 2015. Pharmaceuticals and personal care products in waters: occurrence, toxicity, and risk. Environ. Chem. Lett. 13, 381–394. https://doi.org/10.1007/s10311-015-0524-4

Colborn, T., Vom Saal, F.S., Soto, A.M., 1993. Developmental effects of endocrine-disrupting chemicals in wildlife and humans. Environ. Health Perspect. 101, 378–384. https://doi.org/10.1016/0195-9255(94)90014-0

Corsini, E., Sokooti, M., Galli, C.L., Moretto, A., Colosio, C., 2013. Pesticide induced immunotoxicity in humans: A comprehensive review of the existing evidence. Toxicology 307, 123–135. https://doi.org/10.1016/j.tox.2012.10.009

Coumoul, X., Diry, M., Barouki, R., 2002. PXR-dependent induction of human CYP3A4 gene expression by organochlorine pesticides. Biochem. Pharmacol. 64, 1513–9.

Damalas, C.A., Eleftherohorinos, I.G., 2011. Pesticide exposure, safety issues, and risk assessment indicators. Int. J. Environ. Res. Public Health 8, 1402–1419. https://doi.org/10.3390/ijerph8051402

Davis, A.P., Grondin, C.J., Johnson, R.J., Sciaky, D., King, B.L., McMorran, R., Wiegers, J., Wiegers, T.C., Mattingly, C.J., 2017. The Comparative Toxicogenomics Database: update 2017. Nucleic Acids Res. 45, D972–D978. https://doi.org/10.1093/nar/gkw838

Davis, A.P., Murphy, C.G., Johnson, R., Lay, J.M., Lennon-Hopkins, K., Saraceni-Richards, C., Sciaky, D., King, B.L., Rosenstein, M.C., Wiegers, T.C., Mattingly, C.J., 2013. The Comparative Toxicogenomics Database: update 2013. Nucleic Acids Res. 41, D1104-14. https://doi.org/10.1093/nar/gks994

Deierlein, A.L., Rock, S., Park, S., 2017. Persistent Endocrine-Disrupting Chemicals and Fatty Liver Disease. Curr. Environ. Heal. Reports 4, 439–449. https://doi.org/10.1007/s40572-017-0166-8

Dong, G.-H.H., Tung, K.-Y.Y., Tsai, C.-H.H., Liu, M.-M.M., Wang, D., Liu, W., Jin, Y.-H.H., Hsieh, W.-S.S., Lee, Y.L., Chen, P.-C.C., 2013. Serum polyfluoroalkyl concentrations, asthma outcomes, and immunological markers in a case-control study of Taiwanese children. Environ. Health Perspect. 121, 507–513. https://doi.org/10.1289/ehp.1205351

EFSA, 2008. Polycyclic Aromatic Hydrocarbons in Food Scientific Opinion of the Panel on Contaminants in the Food Chain (Question N° EFSA-Q-2007-136). EFSA J. 724, 1–114.

Fairley, K.J., Purdy, R., Kearns, S., Anderson, S.E., Meade, B.J., 2007. Exposure to the immunosuppressant, perfluorooctanoic acid, enhances the murine IgE and airway hyperreactivity response to ovalbumin. Toxicol. Sci. 97, 375–83. https://doi.org/10.1093/toxsci/kfm053

Fareed, M., Pathak, M.K., Bihari, V., Kamal, R., Srivastava, A.K., Kesavachandran, C.N., 2013. Adverse Respiratory Health and Hematological Alterations among Agricultural Workers Occupationally Exposed to Organophosphate Pesticides: A Cross-Sectional Study in North India. PLoS One 8, e69755. https://doi.org/10.1371/journal.pone.0069755

Fracchiolla, N.S., Annaloro, C., Guidotti, F., Fattizzo, B., Cortelezzi, A., 2016. 2,3,7,8-Tetrachlorodibenzo-p-dioxin (TCDD) role in hematopoiesis and in hematologic diseases: A critical review. Toxicology 374, 60–68. https://doi.org/10.1016/j.tox.2016.10.007

Freeman, L.C., 1978. Centrality in social networks conceptual clarification. Soc. Networks 1, 215–239. https://doi.org/10.1016/0378-8733(78)90021-7

Gallo, V., Leonardi, G., Brayne, C., Armstrong, B., Fletcher, T., 2013. Serum perfluoroalkyl acids concentrations and memory impairment in a large cross-sectional study. BMJ Open 3. https://doi.org/10.1136/bmjopen-2012-002414

Gascon, M., Morales, E., Sunyer, J., Vrijheid, M., 2013. Effects of persistent organic pollutants on the developing respiratory and immune systems: A systematic review. Environ. Int. 52, 51–65. https://doi.org/10.1016/j.envint.2012.11.005

Goh, K.-I., Cusick, M.E., Valle, D., Childs, B., Vidal, M., Barabasi, A.-L., 2007. The human disease network. Proc. Natl. Acad. Sci. 104, 8685–8690. https://doi.org/10.1073/pnas.0701361104

Goldman, S.M., 2014. Environmental Toxins and Parkinson's Disease. Annu. Rev. Pharmacol. Toxicol. 54, 141–164. https://doi.org/10.1146/annurev-pharmtox-011613-135937

Grandjean, P., Clapp, R., 2014. Changing interpretation of human health risks from perfluorinated compounds. Public Health Rep. 129, 482–5. https://doi.org/10.1177/003335491412900605

Gross, L., Birnbaum, L.S., 2017. Regulating toxic chemicals for public and environmental health. PLOS Biol. 15, e2004814. https://doi.org/10.1371/journal.pbio.2004814

Gump, B.B., Wu, Q., Dumas, A.K., Kannan, K., 2011. Perfluorochemical (PFC) Exposure in Children: Associations with Impaired Response Inhibition. Environ. Sci. Technol. 45, 8151–8159. https://doi.org/10.1021/es103712g

Hamed, E.A., Zakhary, M.M., Maximous, D.W., 2012. Apoptosis, angiogenesis, inflammation, and oxidative stress: basic interactions in patients with early and metastatic breast cancer. J. Cancer Res. Clin. Oncol. 138, 999–1009. https://doi.org/10.1007/s00432-012-1176-4

Hernández, A.F., Parrón, T., Tsatsakis, A.M., Requena, M., Alarcón, R., López-Guarnido, O., 2013. Toxic effects of pesticide mixtures at a molecular level: Their

relevance to human health. Toxicology 307, 136–145. https://doi.org/10.1016/j.tox.2012.06.009

Howard, P.H., Muir, D.C.G., 2011. Identifying new persistent and bioaccumulative organics among chemicals in commerce II: pharmaceuticals. Environ. Sci. Technol. 45, 6938–46. https://doi.org/10.1021/es201196x

IARC, 2010. Some non-heterocyclic polycyclic aromatic hydrocarbons and some related exposures. IARC Monogr. Eval. Carcinog. risks to humans 92, 1–853.

Jeong, H., Mason, S.P., Barabási, A.-L., Oltvai, Z.N., 2001. Lethality and centrality in protein networks. Nature 411, 41–42. https://doi.org/10.1038/35075138

Karimi, P., Peters, K.O., Bidad, K., Strickland, P.T., 2015. Polycyclic aromatic hydrocarbons and childhood asthma. Eur. J. Epidemiol. 30, 91–101. https://doi.org/10.1007/s10654-015-9988-6

Khan, M.I., Islam, N., Sahasrabuddhe, A.A., Mahdi, A.A., Siddiqui, H., Ashquin, M., Ahmad, I., 2011. Ubiquitous hazardous metal lead induces TNF-α in human phagocytic THP-1 cells: Primary role of ERK 1/2. J. Hazard. Mater. 189, 255–264. https://doi.org/10.1016/j.jhazmat.2011.02.027

Kim, K.-H., Jahan, S.A., Kabir, E., Brown, R.J.C., 2013. A review of airborne polycyclic aromatic hydrocarbons (PAHs) and their human health effects. Environ. Int. 60, 71–80. https://doi.org/10.1016/j.envint.2013.07.019

Kim, S.-J., Jeong, H.-J., Myung, N.-Y., Kim, M., Lee, J.-H., So, H., Park, R.-K., Kim, H.-M., Um, J.-Y., Hong, S.-H., 2008. The Protective Mechanism of Antioxidants in Cadmium-Induced Ototoxicity in Vitro and in Vivo. Environ. Health Perspect. 116, 854–862. https://doi.org/10.1289/ehp.10467

Knüpfer, H., Preiß, R., 2007. Significance of interleukin-6 (IL-6) in breast cancer (review). Breast Cancer Res. Treat. 102, 129–135. https://doi.org/10.1007/s10549-006-9328-3

Lane, D.P., Cheok, C.F., Lain, S., 2010. p53-based cancer therapy. Cold Spring Harb. Perspect. Biol. 2, a001222. https://doi.org/10.1101/cshperspect.a001222

Law, V., Knox, C., Djoumbou, Y., Jewison, T., Guo, A.C., Liu, Y., Maciejewski, A., Arndt, D., Wilson, M., Neveu, V., Tang, A., Gabriel, G., Ly, C., Adamjee, S., Dame, Z.T., Han, B., Zhou, Y., Wishart, D.S., 2014. DrugBank 4.0: shedding new light on drug metabolism. Nucleic Acids Res. 42, D1091-7. https://doi.org/10.1093/nar/gkt1068

Lindstrom, A.B., Strynar, M.J., Libelo, E.L., 2011. Polyfluorinated Compounds: Past, Present, and Future. Environ. Sci. Technol. 45, 7954–7961. https://doi.org/10.1021/es2011622

Mamane, A., Raherison, C., Tessier, J.-F., Baldi, I., Bouvier, G., 2015. Environmental exposure to pesticides and respiratory health. Eur. Respir. Rev. 24, 462–473. https://doi.org/10.1183/16000617.00006114

Min, Y.-S., Lim, H.-S., Kim, H., 2015. Biomarkers for polycyclic aromatic hydrocarbons and serum liver enzymes. Am. J. Ind. Med. 58, 764–772. https://doi.org/10.1002/ajim.22463

Miralles-Marco, A., Harrad, S., 2015. Perfluorooctane sulfonate: A review of human exposure, biomonitoring and the environmental forensics utility of its chirality and isomer distribution. Environ. Int. 77, 148–159. https://doi.org/10.1016/j.envint.2015.02.002

Nandipati, S., Litvan, I., 2016. Environmental Exposures and Parkinson's Disease. Int. J. Environ. Res. Public Health 13, 881. https://doi.org/10.3390/ijerph13090881

Nicolakakis, N., Hamel, E., 2010. The Nuclear Receptor PPARgamma as a Therapeutic Target for Cerebrovascular and Brain Dysfunction in Alzheimer's Disease. Front. Aging Neurosci. 2. https://doi.org/10.3389/fnagi.2010.00021

Petrulis, J.R., Chen, G., Benn, S., LaMarre, J., Bunce, N.J., 2001. Application of the ethoxyresorufin-O-deethylase (EROD) assay to mixtures of halogenated aromatic compounds. Environ. Toxicol. 16, 177–84.

Reyes-Palomares, A., Rodríguez-López, R., Ranea, J.A.G., Jiménez, F.S., Medina, M.A., 2013. Global Analysis of the Human Pathophenotypic Similarity Gene Network Merges Disease Module Components. PLoS One 8. https://doi.org/10.1371/journal.pone.0056653

Rolland, T., Taşan, M., Charloteaux, B., Pevzner, S.J., Zhong, Q., Sahni, N., Yi, S., Lemmens, I., Fontanillo, C., Mosca, R., Kamburov, A., Ghiassian, S.D., Yang, X., Ghamsari, L., Balcha, D., Begg, B.E., Braun, P., Brehme, M., Broly, M.P., Carvunis, A.-R., Convery-Zupan, D., Corominas, R., Coulombe-Huntington, J., Dann, E., Dreze, M., Dricot, A., Fan, C., Franzosa, E., Gebreab, F., Gutierrez, B.J., Hardy, M.F., Jin, M., Kang, S., Kiros, R., Lin, G.N., Luck, K., MacWilliams, A., Menche, J., Murray, R.R., Palagi, A., Poulin, M.M., Rambout, X., Rasla, J., Reichert, P., Romero, V., Ruyssinck, E., Sahalie, J.M., Scholz, A., Shah, A.A., Sharma, A., Shen, Y., Spirohn, K., Tam, S., Tejeda, A.O., Trigg, S.A., Twizere, J.-C., Vega, K., Walsh, J., Cusick, M.E., Xia, Y., Barabási, A.-L., Iakoucheva, L.M., Aloy, P., De Las Rivas, J., Tavernier, J., Calderwood, M.A., Hill, D.E., Hao, T., Roth, F.P., Vidal, M., 2014. A proteome-scale map of the human interactome network. Cell 159, 1213–1226. https://doi.org/10.1016/j.cell.2014.10.050.A

Rosen, M.B., Das, K.P., Rooney, J., Abbott, B., Lau, C., Corton, J.C., 2017. PPARα-independent transcriptional targets of perfluoroalkyl acids revealed by transcript profiling. Toxicology 387, 95–107. https://doi.org/10.1016/j.tox.2017.05.013

Rota, M., Bosetti, C., Boccia, S., Boffetta, P., La Vecchia, C., 2014. Occupational exposures to polycyclic aromatic hydrocarbons and respiratory and urinary tract cancers: an updated systematic review and a meta-analysis to 2014. Arch. Toxicol. 88, 1479–1490. https://doi.org/10.1007/s00204-014-1296-5

Rual, J.-F., Venkatesan, K., Hao, T., Hirozane-Kishikawa, T., Dricot, A., Li, N., Berriz, G.F., Gibbons, F.D., Dreze, M., Ayivi-Guedehoussou, N., Klitgord, N., Simon, C., Boxem, M., Milstein, S., Rosenberg, J., Goldberg, D.S., Zhang, L. V., Wong, S.L., Franklin, G., Li, S., Albala, J.S., Lim, J., Fraughton, C., Llamosas, E., Cevik, S., Bex, C., Lamesch, P., Sikorski, R.S., Vandenhaute, J., Zoghbi, H.Y., Smolyar, A., Bosak, S., Sequerra, R., Doucette-Stamm, L., Cusick, M.E., Hill, D.E., Roth, F.P., Vidal, M., 2005. Towards a proteome-scale map of the human protein–protein interaction network. Nature 437, 1173–1178. https://doi.org/10.1038/nature04209

Sánchez-Santed, F., Colomina, M.T., Herrero Hernández, E., 2016. Organophosphate pesticide exposure and neurodegeneration. Cortex 74, 417–426. https://doi.org/10.1016/j.cortex.2015.10.003

Silins, I., Högberg, J., 2011. Combined toxic exposures and human health: biomarkers of exposure and effect. Int. J. Environ. Res. Public Health 8, 629–47. https://doi.org/10.3390/ijerph8030629

Steenland, K., Fletcher, T., Savitz, D.A., 2010. Epidemiologic Evidence on the Health Effects of Perfluorooctanoic Acid (PFOA). Environ. Health Perspect. 118, 1100–1108. https://doi.org/10.1289/ehp.0901827

Stelzl, U., Worm, U., Lalowski, M., Haenig, C., Brembeck, F.H., Goehler, H., Stroedicke, M., Zenkner, M., Schoenherr, A., Koeppen, S., Timm, J., Mintzlaff, S., Abraham, C., Bock, N., Kietzmann, S., Goedde, A., Toksöz, E., Droege, A., Krobitsch, S., Korn, B., Birchmeier, W., Lehrach, H., Wanker, E.E., 2005. A Human Protein-Protein Interaction Network: A Resource for Annotating the Proteome. Cell 122, 957–968. https://doi.org/10.1016/j.cell.2005.08.029

Sturgill, M.G., Lambert, G.H., 1997. Xenobiotic-induced hepatotoxicity: mechanisms of liver injury and methods of monitoring hepatic function. Clin. Chem. 43, 1512–26.

Takemoto, K., Oosawa, C., 2012. Introduction to Complex Networks: Measures, Statistical Properties, and Models, in: Statistical and Machine Learning Approaches for Network Analysis. John Wiley & Sons, Inc., Hoboken, NJ, USA, pp. 45–75. https://doi.org/10.1002/9781118346990.ch2

Tanoue, R., Nomiyama, K., Nakamura, H., Kim, J.-W., Isobe, T., Shinohara, R., Kunisue, T., Tanabe, S., 2015. Uptake and Tissue Distribution of Pharmaceuticals and Personal Care Products in Wild Fish from Treated-Wastewater-Impacted Streams. Environ. Sci. Technol. 49, 11649–58. https://doi.org/10.1021/acs.est.5b02478

Tao, G.-H., Yang, L.-Q., Gong, C.-M., Huang, H.-Y., Liu, J.-J.J.-D., Liu, J.-J.J.-D., Yuan, J.-H., Chen, W., Zhuang, Z.-X., 2009. Effect of PARP-1 deficiency on DNA damage and repair in human bronchial epithelial cells exposed to Benzo(a)pyrene. Mol. Biol. Rep. 36, 2413–2422. https://doi.org/10.1007/s11033-009-9472-z

Tavakoly Sany, S.B., Hashim, R., Salleh, A., Rezayi, M., Karlen, D.J., Razavizadeh, B.B.M., Abouzari-lotf, E., 2015. Dioxin risk assessment: mechanisms of action and possible toxicity in human health. Environ. Sci. Pollut. Res. 22, 19434–19450. https://doi.org/10.1007/s11356-015-5597-x

Tchounwou, P.B., Yedjou, C.G., Patlolla, A.K., Sutton, D.J., 2012. Molecular, Clinical and Environmental Toxicology 101, 1–30. https://doi.org/10.1007/978-3-7643-8340-4

UNEP, 2009. STOCKHOLM CONVENTION ON PERSISTENT ORGANIC POLLUTANTS (POPs) Text and Annexes.

US EPA, 2014. Appendix A to 40 CFR, Part 423.

US EPA, 2004. Exposure and Human Health Reassessment of 2,3,7,8-Tetrachlorodibenzo-P-Dioxin (Tcdd) and Related Compounds National Academy Sciences (External Review Draft). Washington, D.C.

Veldhoen, M., Hirota, K., Westendorf, A.M., Buer, J., Dumoutier, L., Renauld, J.-C., Stockinger, B., 2008. The aryl hydrocarbon receptor links TH17-cell-mediated autoimmunity to environmental toxins. Nature 453, 106–109. https://doi.org/10.1038/nature06881

Vinggaard, A.M., Hnida, C., Larsen, J.C., 2000. Environmental polycyclic aromatic hydrocarbons affect androgen receptor activation in vitro. Toxicology 145, 173–83.

VoPham, T., Bertrand, K.A., Hart, J.E., Laden, F., Brooks, M.M., Yuan, J.-M., Talbott, E.O., Ruddell, D., Chang, C.-C.H., Weissfeld, J.L., 2017. Pesticide exposure and liver cancer: a review. Cancer Causes Control 28, 177–190. https://doi.org/10.1007/s10552-017-0854-6

WHO and UNEP, 2013. State of the science of endocrine disrupting chemicals - 2012 : an assessment of the state of the science of endocrine disruptors prepared by a group of experts for the United Nations Environment Programme (UNEP) and WHO. United National Environment Programme.

Wolf, C.J., Rider, C. V., Lau, C., Abbott, B.D., 2014. Evaluating the additivity of perfluoroalkyl acids in binary combinations on peroxisome proliferator-activated receptor-α activation. Toxicology 316, 43–54. https://doi.org/10.1016/j.tox.2013.12.002

Ye, M., Beach, J., Martin, J., Senthilselvan, A., 2013. Occupational Pesticide Exposures and Respiratory Health. Int. J. Environ. Res. Public Health 10, 6442–6471. https://doi.org/10.3390/ijerph10126442

Yıldırım, M.A., Goh, K.-I., Cusick, M.E., Barabási, A.-L., Vidal, M., 2007. Drug—target network. Nat. Biotechnol. 25, 1119–1126. https://doi.org/10.1038/nbt1338

## Figure legends

**Figure 1.** Partial chemical-target network. Circles and rectangles correspond to chemicals (environmental contaminants) and target proteins, respectively. Network was extracted from the whole chemical-target network by selecting nodes with a degree > 20. Node size indicates node degree. Chemical nodes (circles) are colored according to chemical categories. Network layout (e.g., position of nodes and lengths of edges) is based on the prefuse force directed layout algorithm in Cytoscape (version 3.5.1). Abbreviations of chemicals represented here are: BDE49, 2,2′,4,5′-tetrabromodiphenyl ether; HBCD, hexabromocyclododecane; NiSO$_4$, nickel(II) sulfate. Abbreviations for genes are; BCL2, B-cell lymphoma 2; CAT, Catalase; HIF1A, hypoxia-inducible factor 1 alpha; HMOX1, heme oxygenase 1; IL, interleukin; MAPK, mitogen-activated protein kinase; NR1I3, nuclear receptor subfamily 1 group I member 3; PGR, progesterone receptor; PTGS2, prostaglandin-endoperoxide synthase; RELA, transcription factor p65.

**Figure 2.** Degree distributions of chemical-target networks. Distribution of number of (a) genes targeted by a chemical ($n_c$) and (b) chemicals targeting a gene ($n_g$). CAR, constitutive androstane receptor; PTPN11, protein tyrosine phosphatase, non-receptor type 11; SLC22A2 and SLC22A11, solute carrier family 22 member 2 and member 11.

**Figure 3.** Boxplot of node degree [log2(degree)] of chemical targets in human protein–protein interactions (PPI) network of each chemical category. See Sec. 2 for details of chemical categories. Proteins in the *multiple* category are targeted by chemical(s) in multiple chemical categories. Node degrees of any (*all*) and essential proteins (*ESGs*) also displayed for comparison; *p*-value, using Wilcoxon rank-sum test.

**Figure 4.** Ratio of essential proteins to target proteins for each chemical category. See Figure 3 for category definitions; *p*-value, using Fisher's exact test.

**Figure 5.** Distribution of distances between chemical targets and disease-associated proteins (red). Distribution of random controls (i.e., distances between randomly selected and disease-associated proteins) also shown (blue); *p*-value, using the Kolmogorov-Smirnov (KS) test.

Figure 6. Difference in mean distance between actual data ($D_{real}$) and random control ($D_{rand}$) values evaluated using Welch's two-sample t-test.

**Figure S1.** Partial human protein–protein interaction network. Only targets of *dioxins* and disease-associated proteins are presented. Node size corresponds to node degree. Rectangular nodes indicate chemical (*dioxins*) targets; circle nodes indicate genes not targeted by chemical (*dioxins*). Nodes are colored based on related disease classes. Rectangle nodes with red edge indicate that proteins are both chemical targets and disease-associated.

**Figure S2.** Boxplots for centrality indicators of chemical targets (proteins) in human protein–protein interactions (PPI) network according to chemical categories. (a) Betweenness centrality [log2(betweenness)], (b) closeness centrality [log2(closeness)], and (c) clustering coefficient centrality [log2(clustering coefficient)] were calculated for each node. See Figure 3 for category definitions; *p*-value, using Wilcoxon rank-sum test. ESGs, essential proteins.

**Figure S3.** Difference in mean distance between actual data ($D_{real}$) and a random control ($D_{rand}$) for 14 disease categories; *p*-value, using Welch's two-sample *t*-test.

**Table S1.** List of nodes for human protein–protein interactions (PPI). Each row indicates interaction pair.

**Table S2.** Chemicals and corresponding categories, and target genes obtained from Comparative Toxicogenomics Database (CTD, as of July 06, 2017). Numbers (0/1) indicate whether chemical belongs to category (1) or not (0).

**Table S3.** Network properties of nodes in chemical-target network

**Table S4.** List of top five hub chemicals in chemical-target network
Values in brackets indicate number of genes targeted by the chemical.

**Table S5.** List of top five hub target proteins in chemical-target network
Values in brackets indicate number of chemicals targeting the gene.

**Table S6.** List of *p*-values of differences between $D_{real}$ and $D_{rand}$

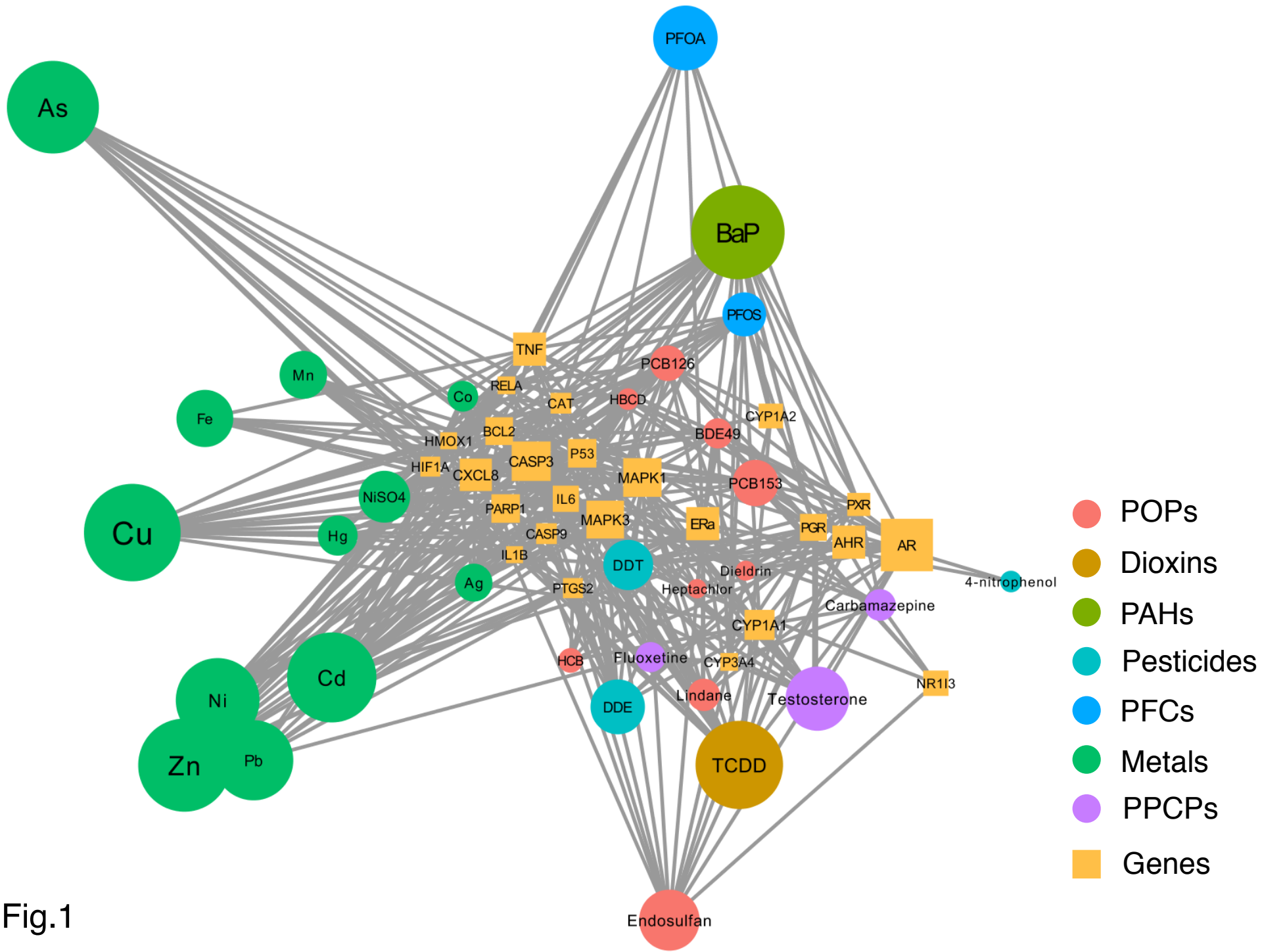

Fig.1

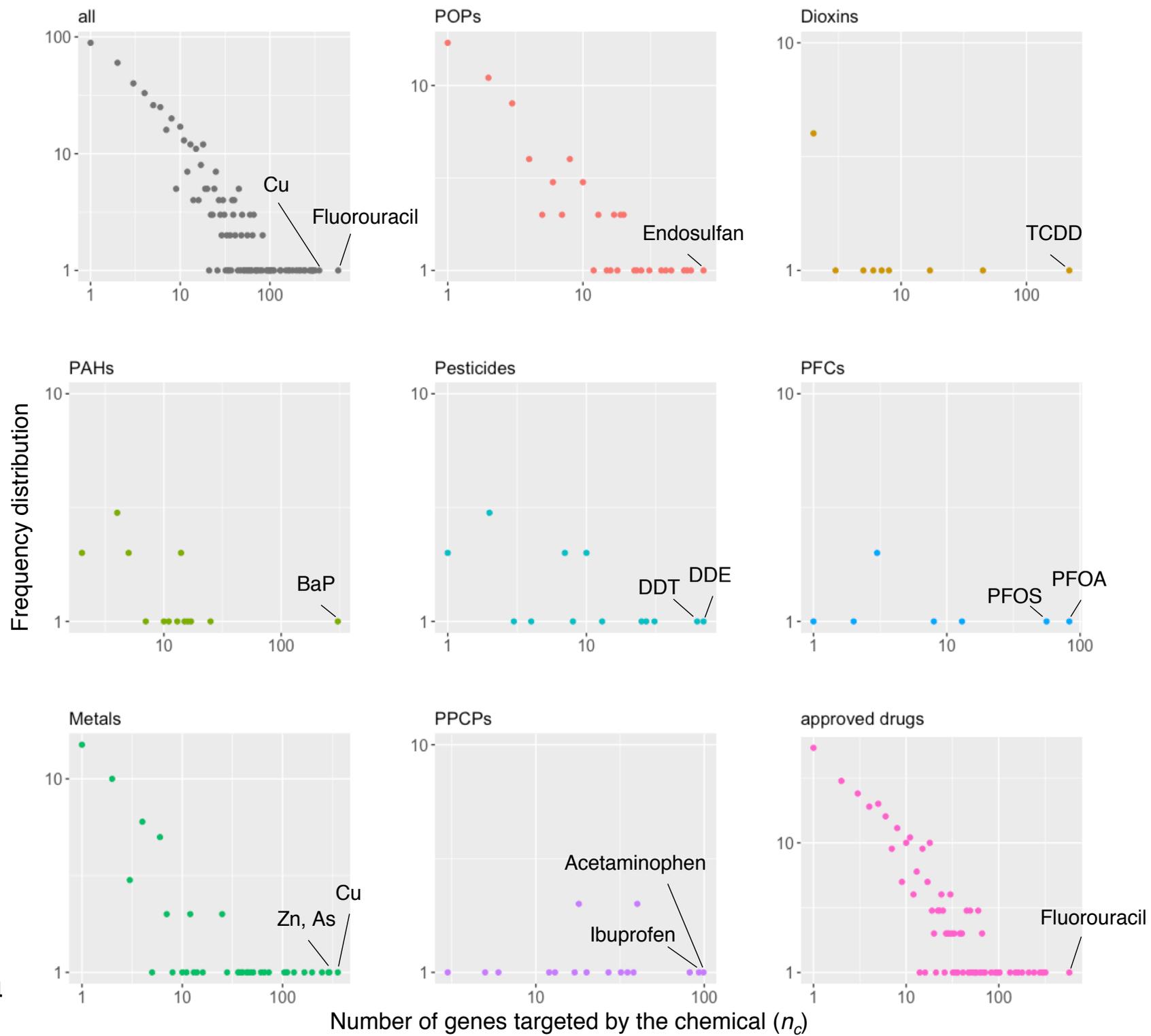

Fig. 2a

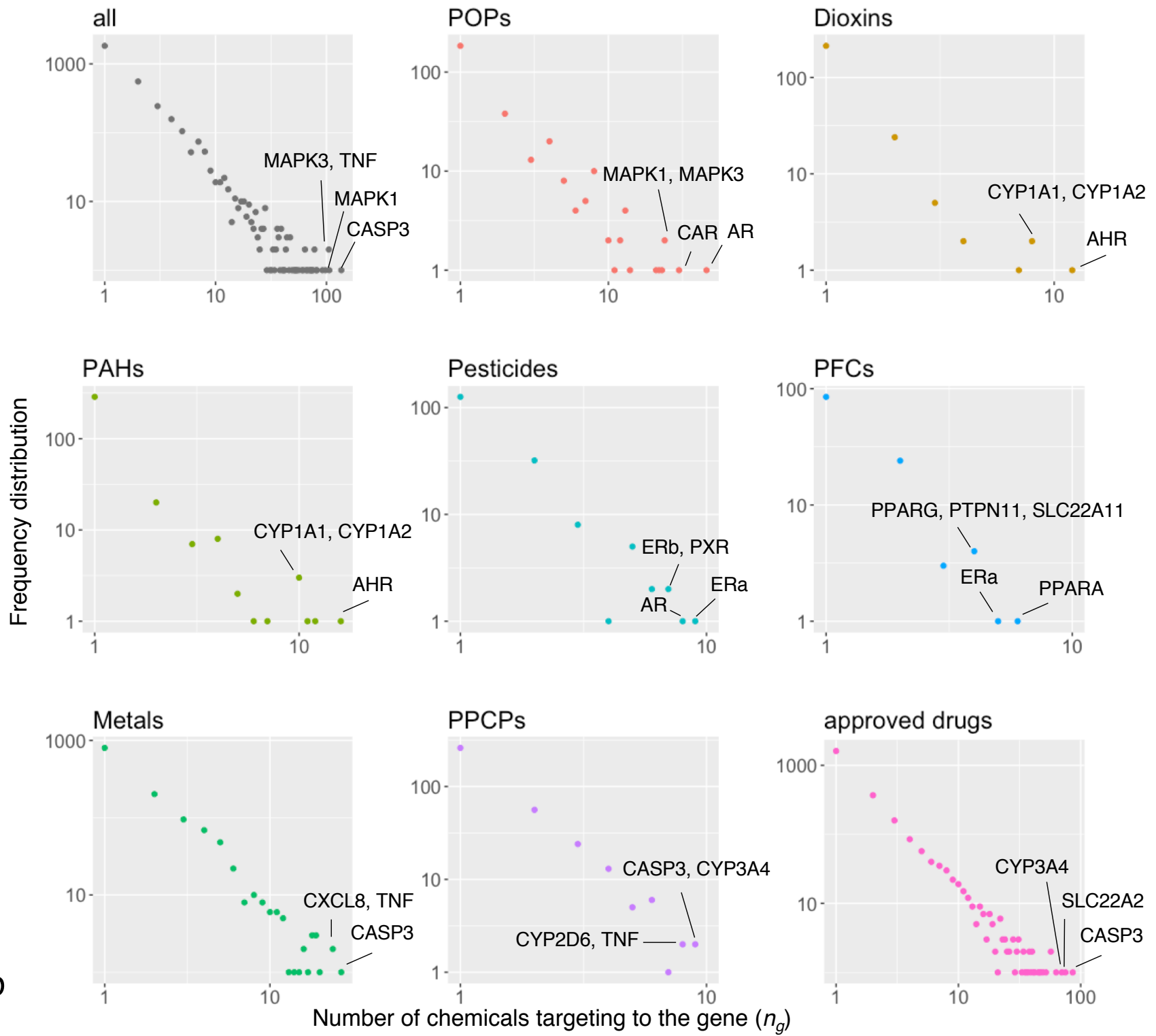

Fig. 2b

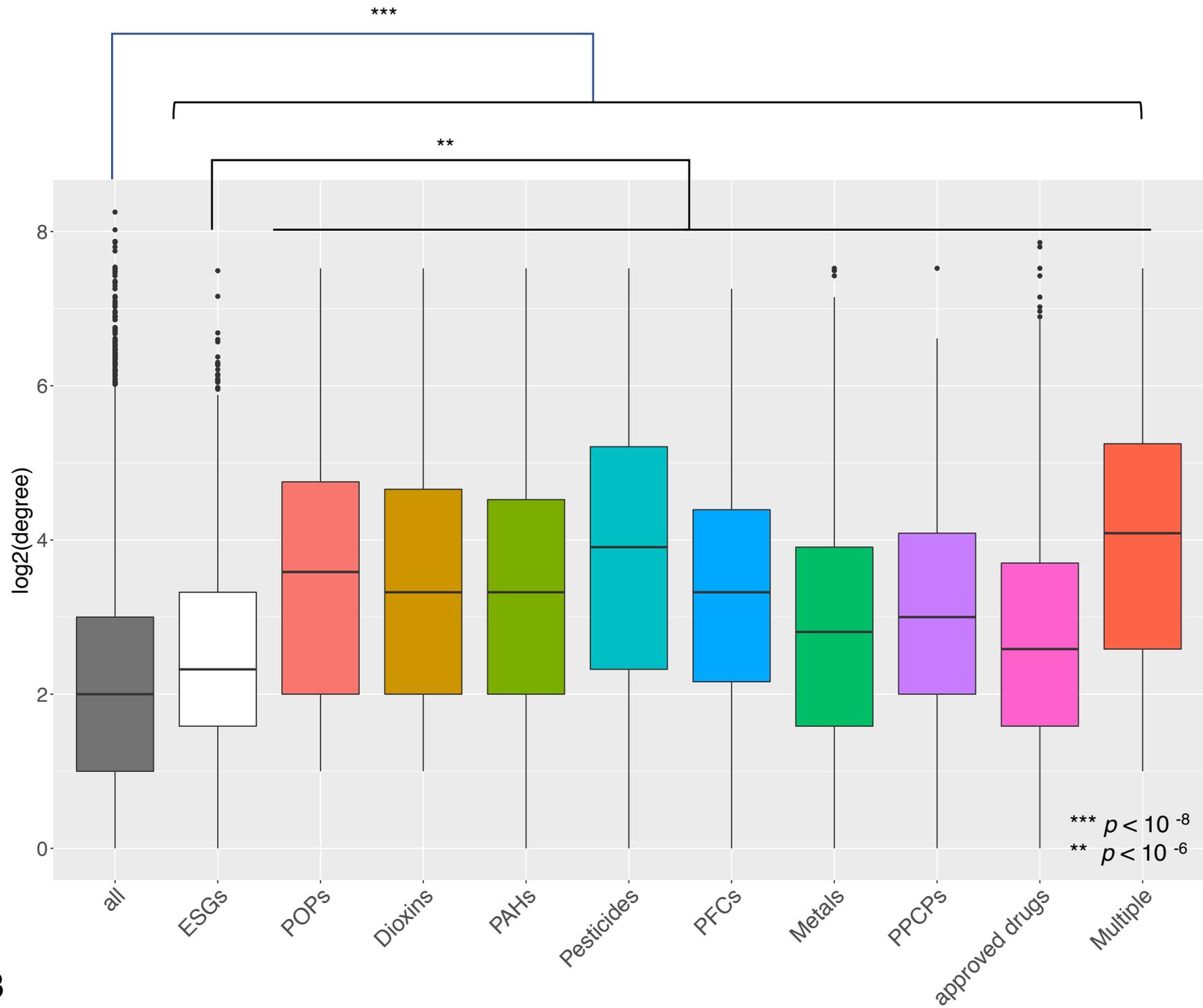

Fig. 3

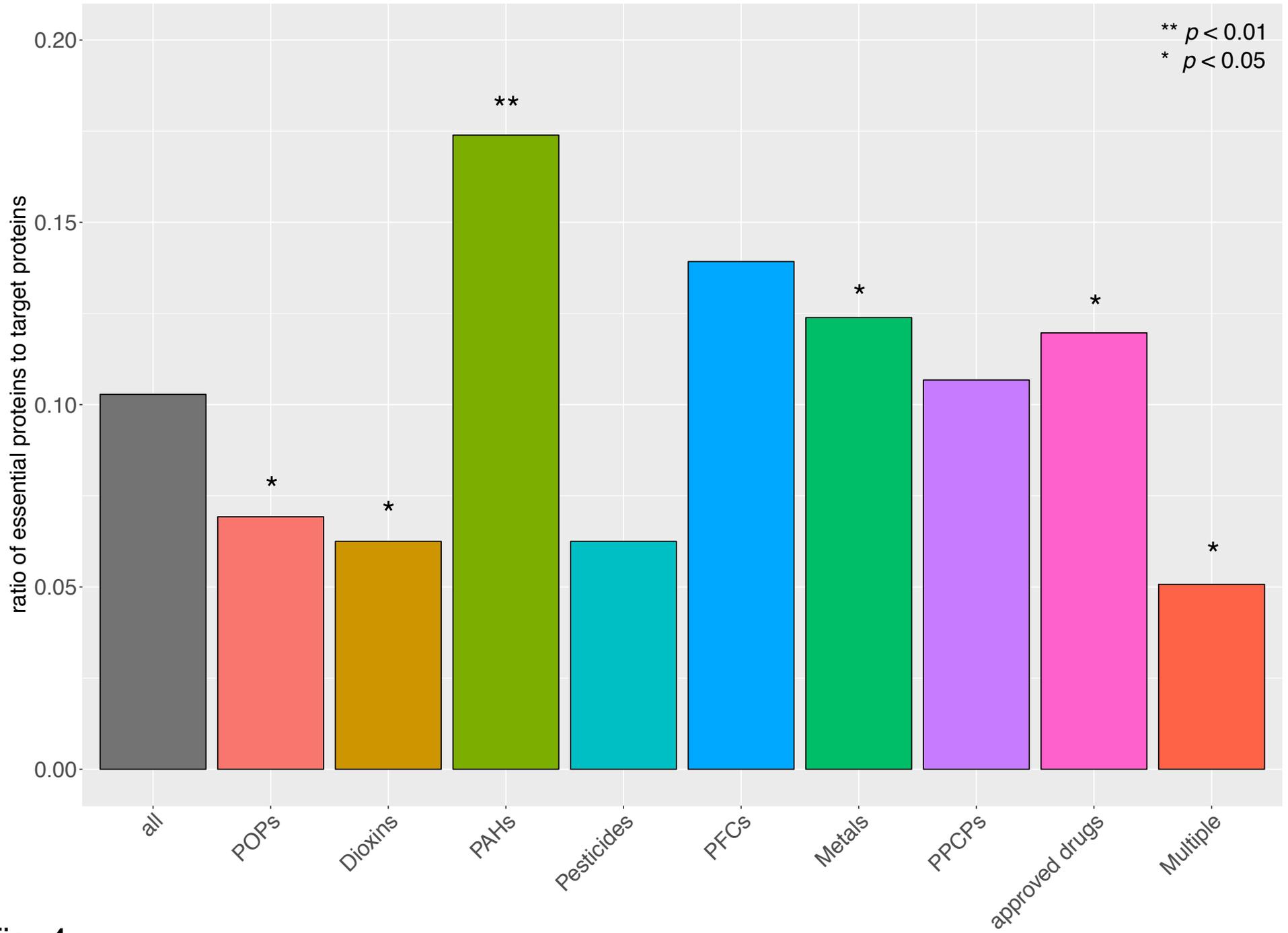

Fig. 4

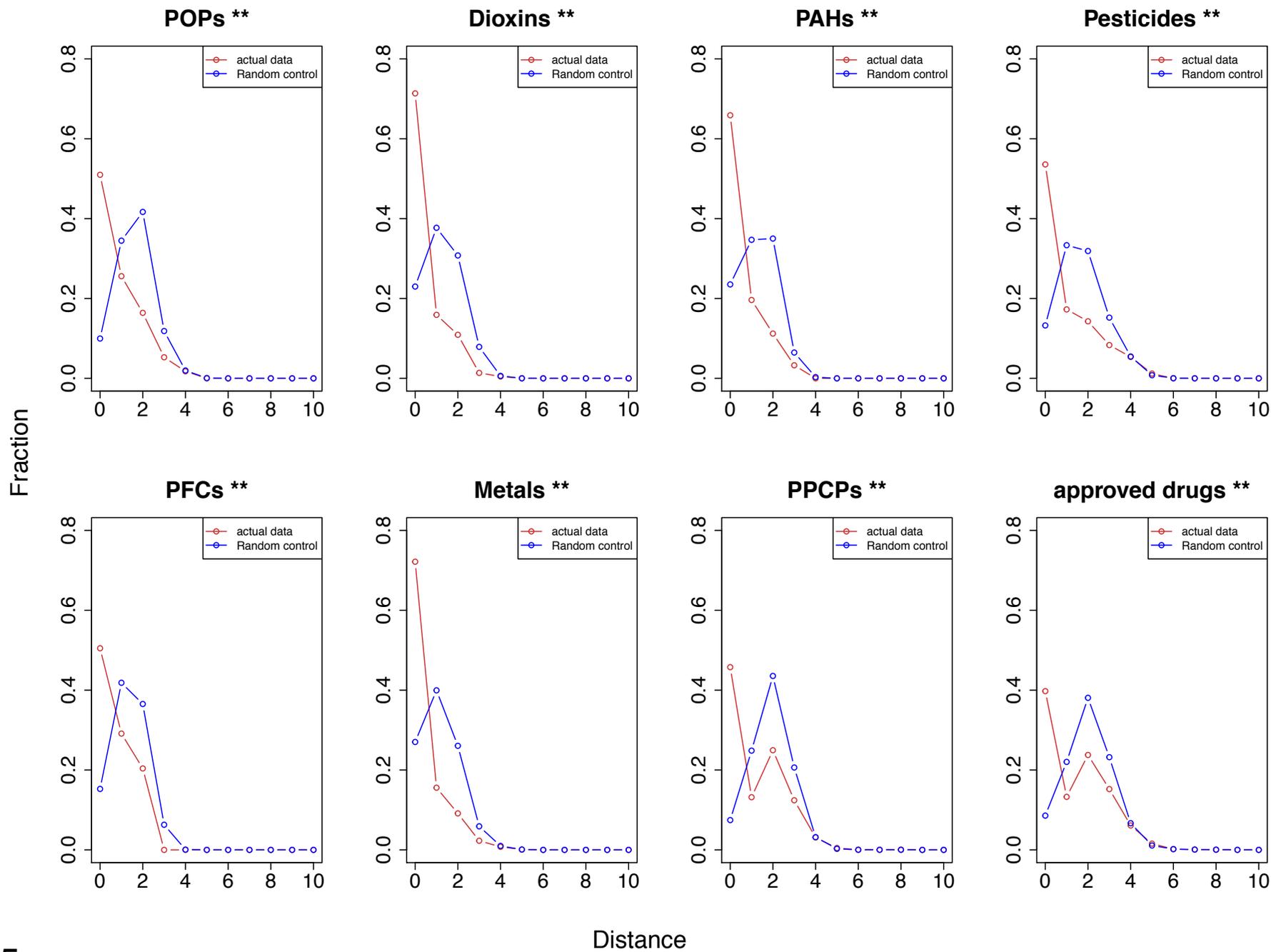

Fig.5

**p < 0.01

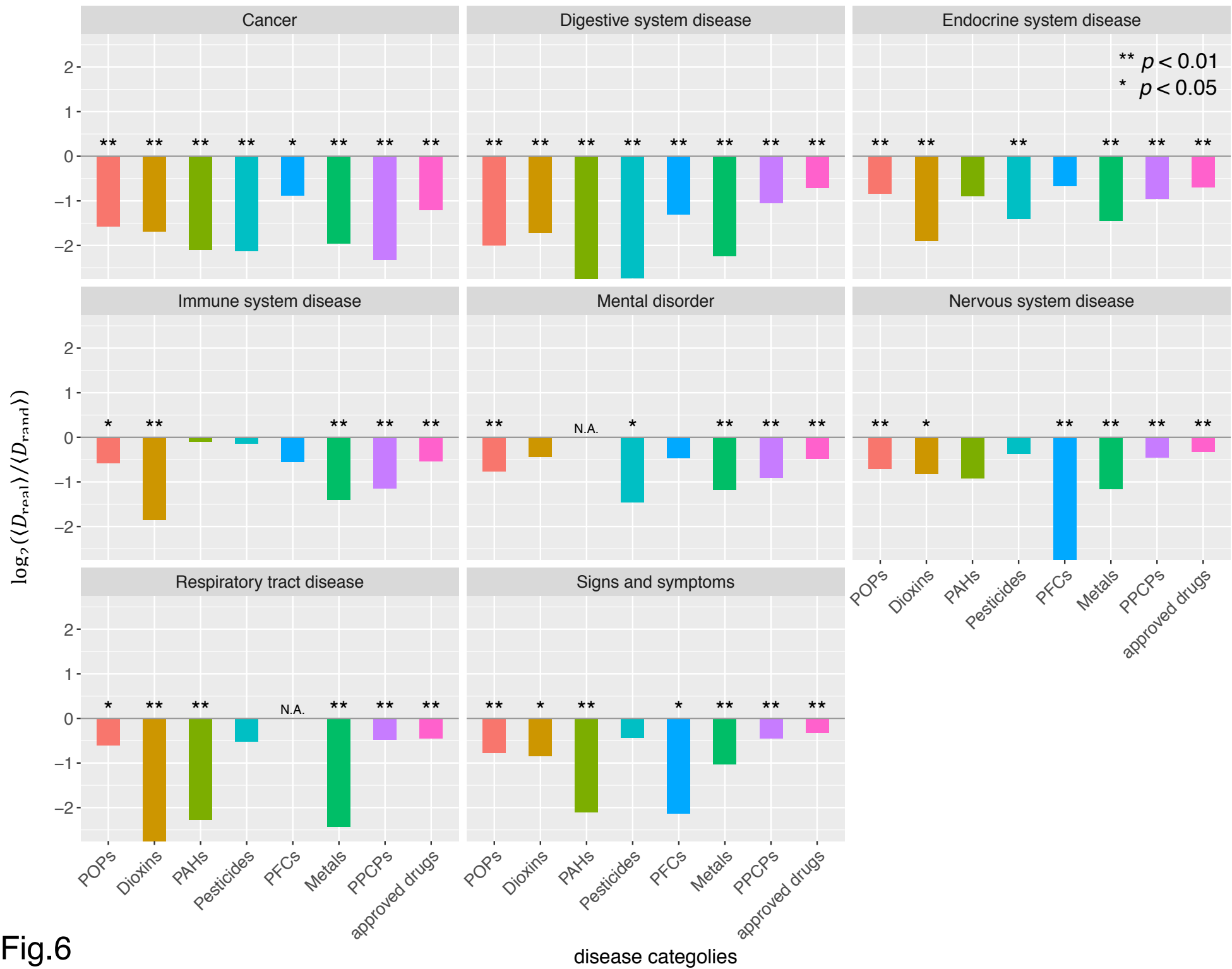

Fig.6

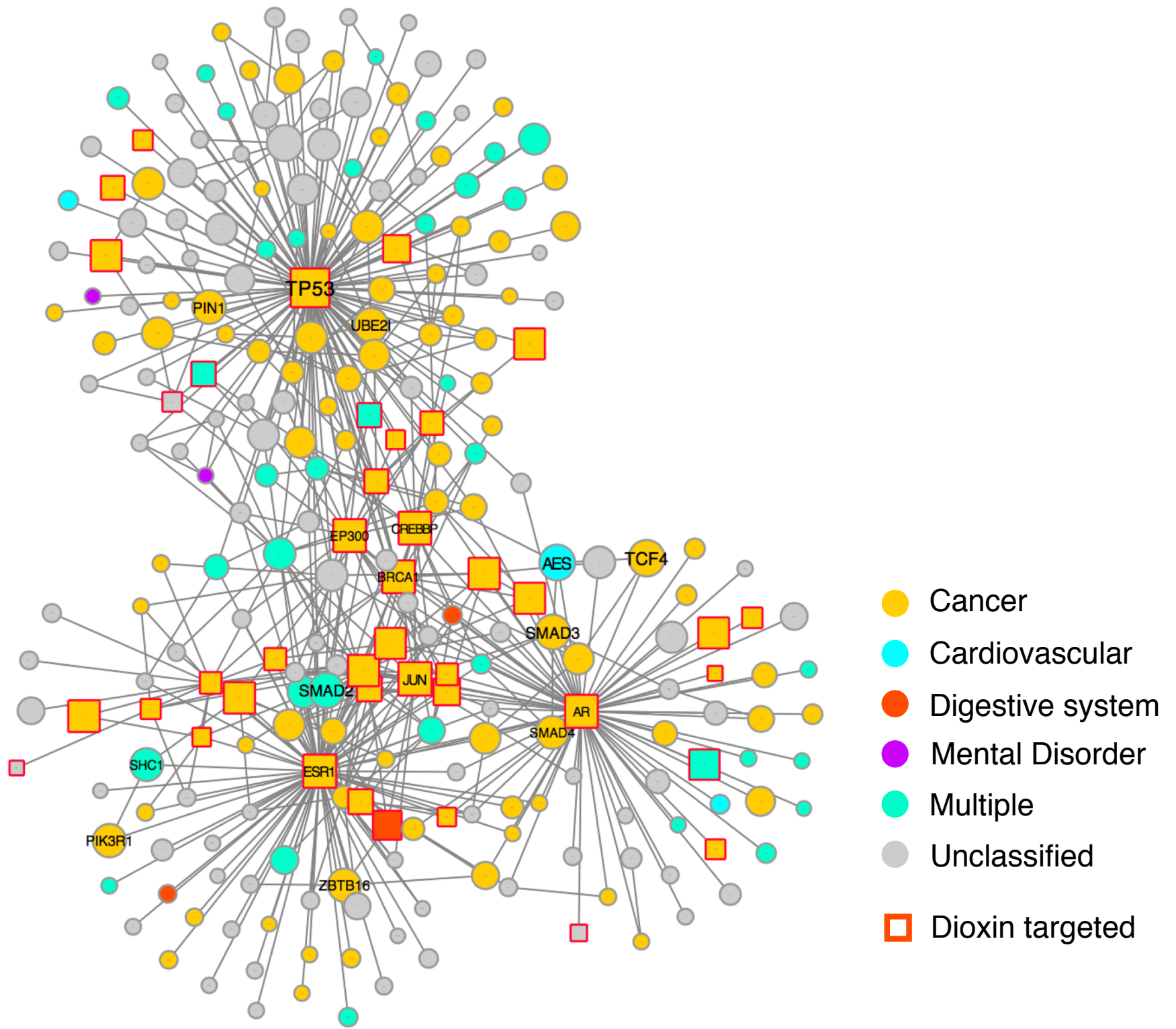

Fig. S1

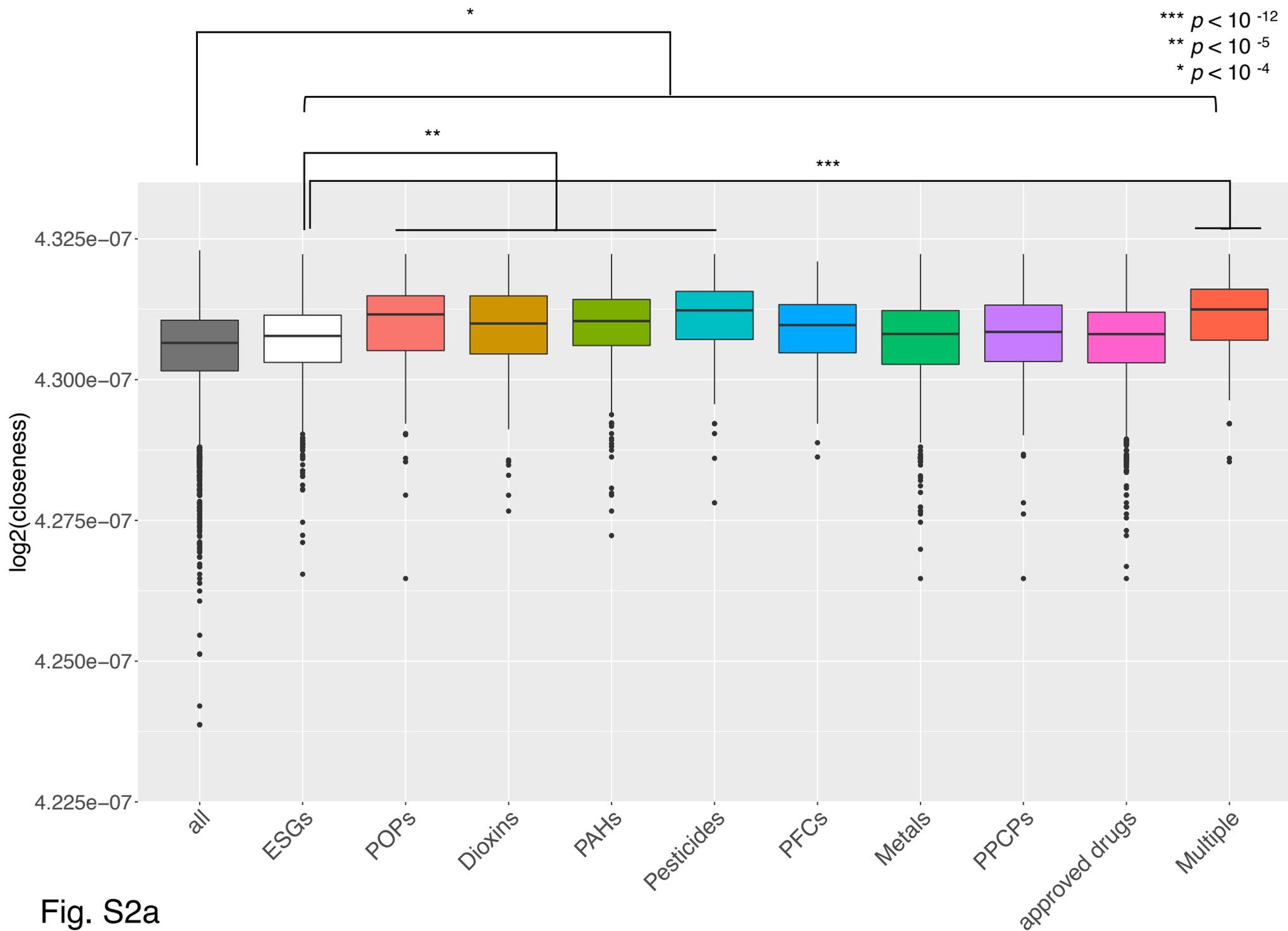

Fig. S2a

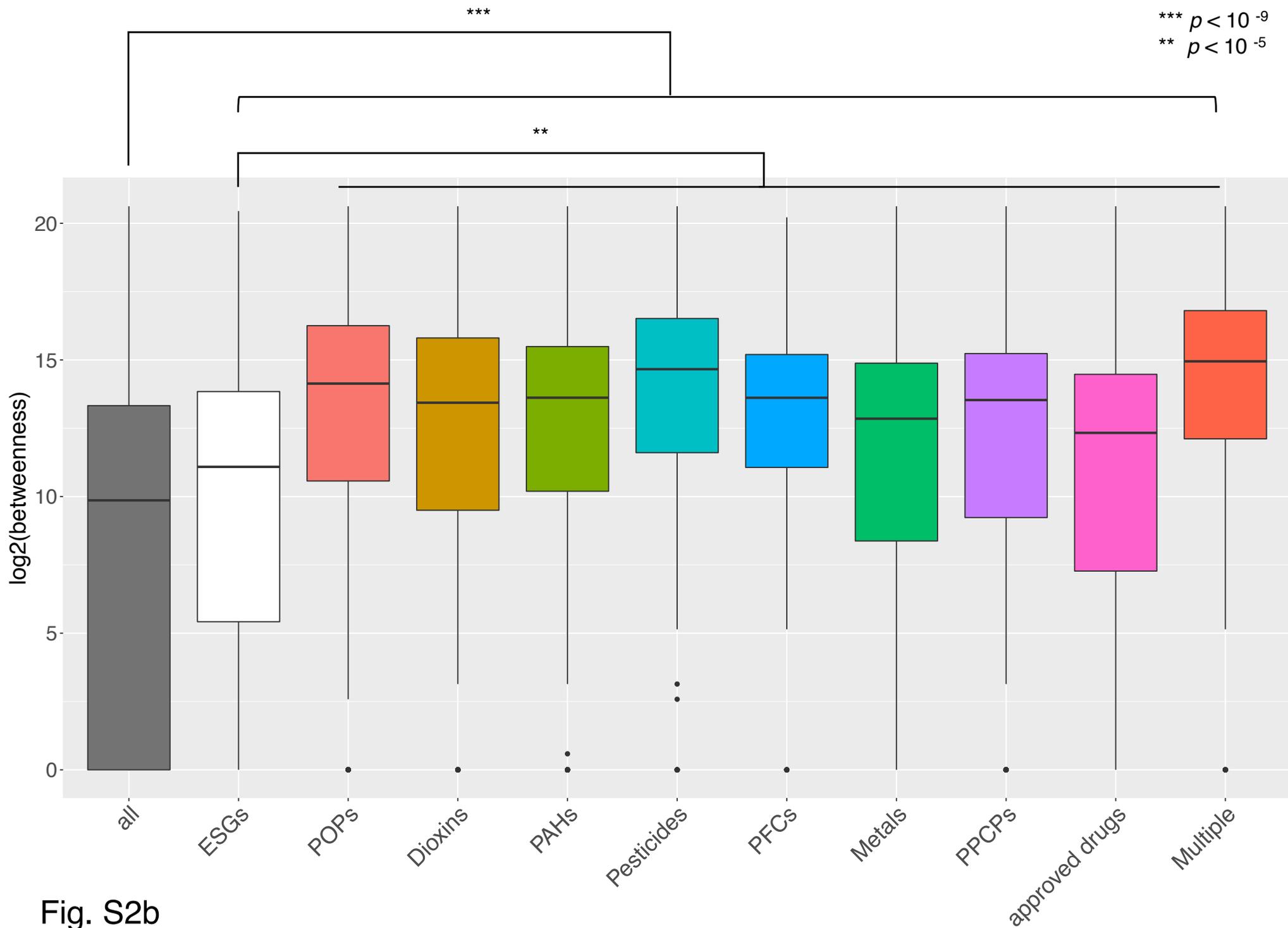

Fig. S2b

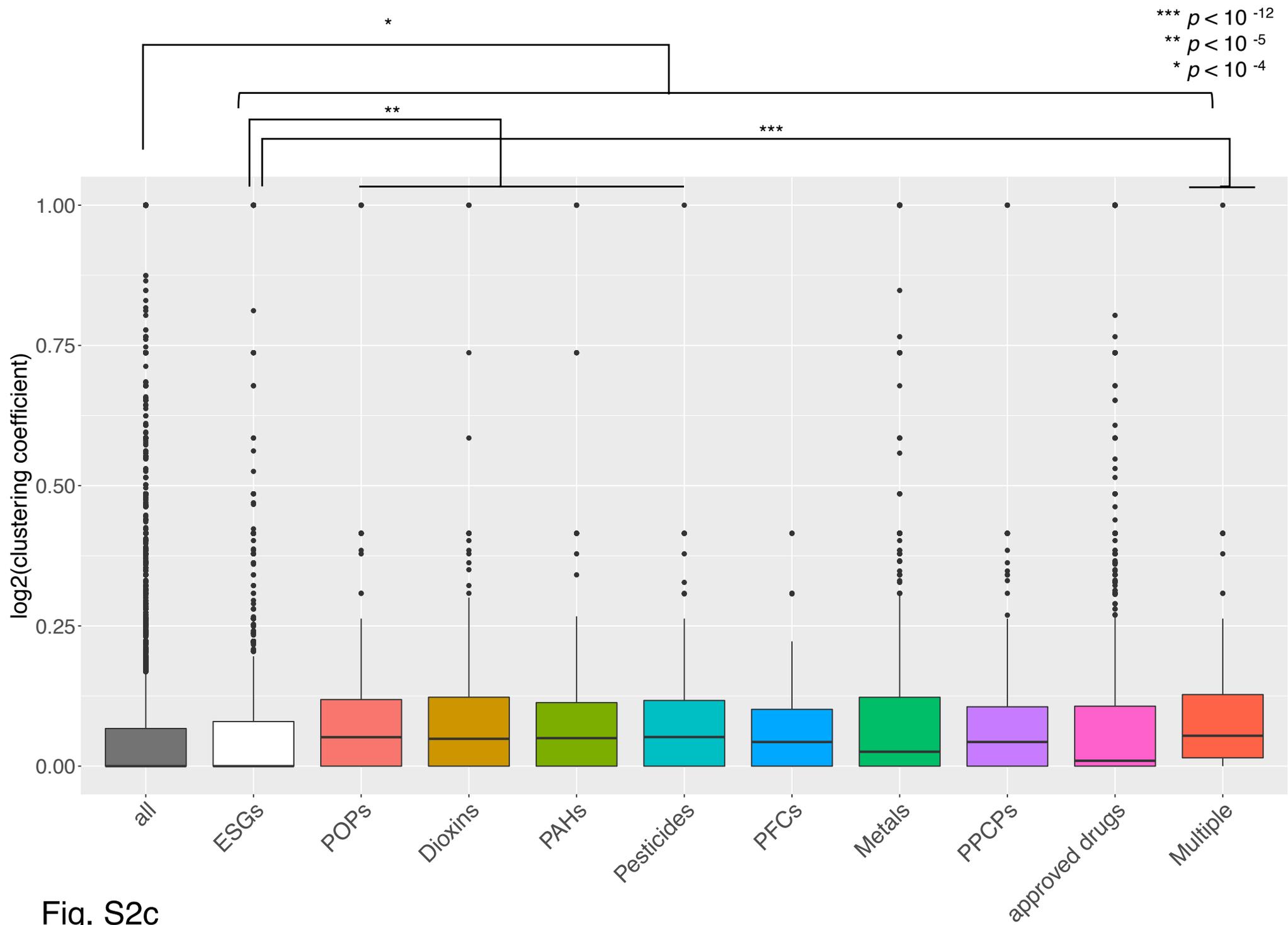

Fig. S2c

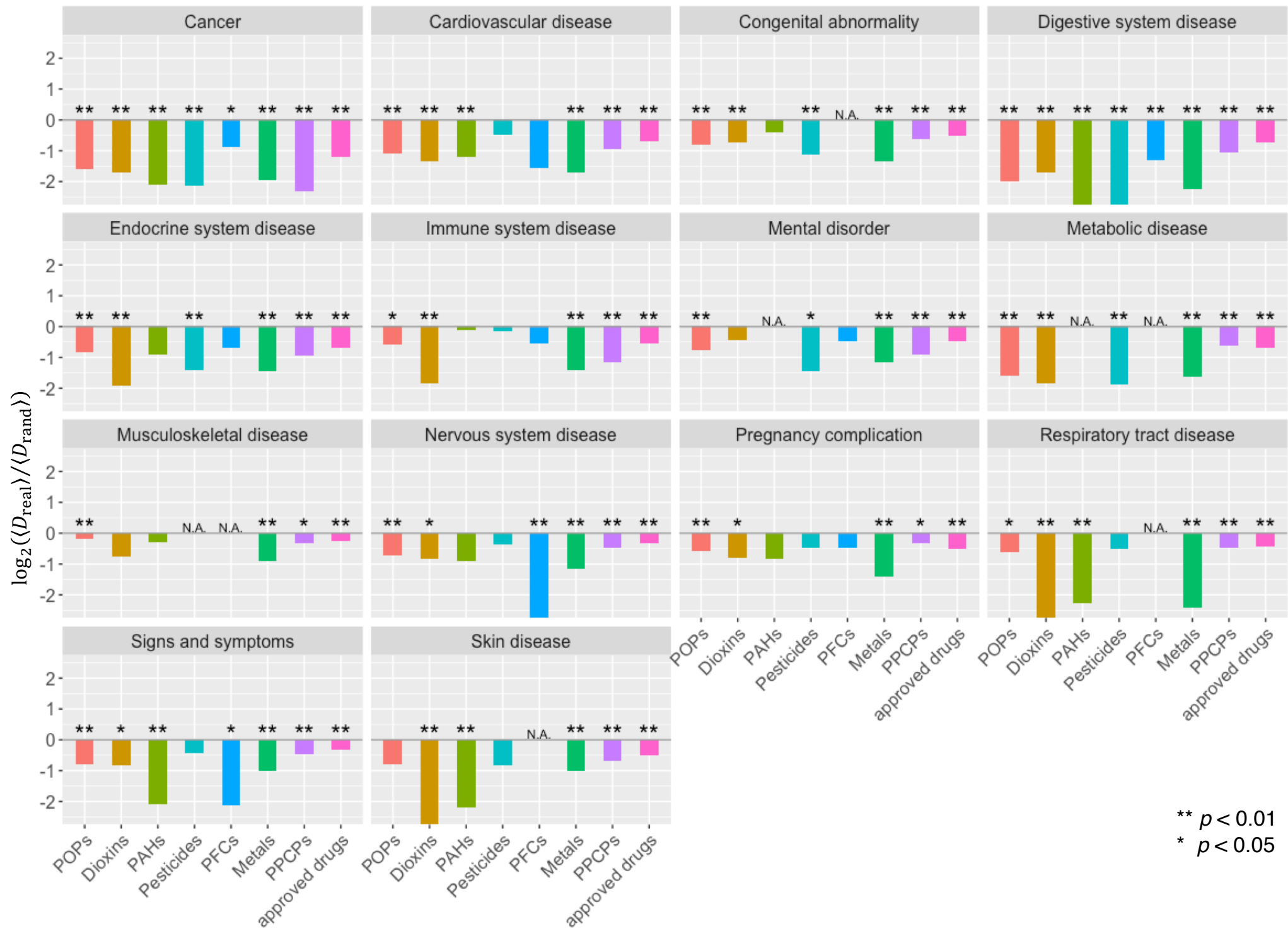

Fig.S3